\documentclass{ws-ijmpa}

\def\be{\begin{eqnarray}}
\def\ee{\end{eqnarray}}
\def\bea{\begin{eqnarray}}
\def\eea{\end{eqnarray}}
\def\beas{\begin{eqnarray*}}
\def\eeas{\end{eqnarray*}}

\begin{document}

\markboth{Matthias Burkardt}
{Impact Parameter Space Interpretation for
Generalized Parton Distributions}

\catchline{}{}{}

\title{IMPACT PARAMETER SPACE INTERPRETATION
FOR GENERALIZED PARTON DISTRIBUTIONS}

\author{\footnotesize MATTHIAS BURKARDT\footnote{
Typeset names in 8 pt roman, uppercase. Use the footnote to indicate the
present or permanent address of the author.}}

\address{Department of Physics,
New Mexico State University\\
Las Cruces, New Mexico 88011, U.S.A.
\footnote{State completely without abbreviations, the
affiliation and mailing address, including country. Typeset in 8 pt
italic.}
}

\maketitle

\pub{Received (Day Month Year)}{Revised (Day Month Year)}

\begin{abstract}
The Fourier transform of generalized parton 
distribution functions at $\xi=0$ describes the
distribution of partons in the transverse plane.
The physical significance of these impact 
parameter dependent parton distribution 
functions is discussed. In particular,
it is shown that they satisfy positivity
constraints which justify their physical
interpretation as a probability density.
The generalized parton distribution $H$ is related
to the impact parameter distribution of
unpolarized quarks for an unpolarized nucleon,
$\tilde{H}$ is related to the distribution
of longitudinally polarized quarks in a
longitudinally polarized nucleon, and $E$ is
related to the distortion of the unpolarized
quark distribution in the transverse plane when
the nucleon has transverse polarization. 
The magnitude of the resulting transverse
flavor dipole moment can be related to the
anomalous magnetic moment for that flavor in
a model independent way.
\end{abstract}
\section{Introduction}
Deep-inelastic scattering (DIS) experiments have not
only played a crucial role in establishing QCD as a
theory of strong interactions but have also been an 
important tool for
exploring the quark-gluon structure of hadrons. In
the Bjorken scaling limit these experiments allow
probing parton distribution functions (PDFs), 
which have a very physical interpretation as the 
(probability) density for finding partons carrying 
the fraction $x$ of the target's total momentum in 
the infinite momentum (or light-cone) frame. 
In particular, due to the probabilistic 
interpretation of PDFs, results from DIS 
experiments have contributed substantially to our
present intuitive understanding of the quark gluon 
structure of hadrons. 

PDFs are defined as the forward matrix element of 
a light-like correlation function, i.e.
\be 
q(x)
&=&\int\!\! \frac{dx^-}{4\pi}\langle P,S|
\bar{q}(-\frac{x^-}{2},{\bf 0}_\perp)
\gamma^+ q(\frac{x^-}{2},{\bf 0}_\perp)
|P,S\rangle e^{ix{p}^+x^-},
\nonumber\\ 
\Delta q(x)S^+ &=& P^+
\int\!\! \frac{dx^-}{4\pi}\langle P,S|
\bar{q}(-\frac{x^-}{2},{\bf 0}_\perp)
\gamma^+ \gamma_5 q(\frac{x^-}{2},{\bf 0}_\perp)
|P,S\rangle e^{ix{p}^+x^-}.
\nonumber\\
\label{eq:pd}
\ee
Throughout this work, we will use light-cone gauge 
$A^+=0$. In all other gauges, a straight line gauge 
string connecting the quark field operators needs to
be included in this definition (\ref{eq:pd}).
When sandwiched between states with the same 
light-cone momentum $p^+$, the operator 
\be
\hat{O}_q(x,{\bf 0}_\perp) 
\equiv \int\!\! \frac{dx^-}{4\pi}
\bar{q}(-\frac{x^-}{2},{\bf 0}_\perp)
\gamma^+ q(\frac{x^-}{2},{\bf 0}_\perp)
e^{ix{p}^+x^-}
\label{eq:0perp}
\ee
in Eq. (\ref{eq:pd})
has the effect of filtering out (the light-cone 
`good' component of) quark fields with momentum 
fraction $x$, which is why PDFs `measure' 
the light-cone momentum density of the quarks.

More recently, generalized parton distributions 
(GPDs) \cite{Mue}
have attracted a considerable amount of 
interest (for a recent review, see for example Ref. 
\cite{vdh}), since it has been pointed out that on 
the one hand GPDs can be related to the total angular
momentum carried by quarks in the nucleon and on the
other hand they play an important role in deeply
virtual Compton scattering experiments \cite{xdj}.
GPDs are defined as
matrix elements of the same operator that is used
to define conventional PDFs, except that GPDs are
defined as transition matrix elements between states
with different momenta (and perhaps also different
spins)
\be
& &\int\!\! \frac{dx^-}{4\pi}\langle P^\prime,S^\prime|
\bar{q}(-\frac{x^-}{2},{\bf 0_\perp})
\gamma^+ q(\frac{x^-}{2},{\bf 0}_\perp)
|P,S\rangle 
e^{ix\bar{p}^+x^-}
\nonumber\\
& &\quad \quad= \frac{1}{2\bar{p}^+}
\bar{u}(p^\prime,s^\prime)\left(\gamma^+  
H_q(x,\xi,t)
+ i\frac{\sigma^{+\nu}\Delta_\nu}{2M} E_q(x,\xi,t)
\right)u(p,s) \label{eq:gpd}
\\[1.ex]
& &
\int\!\! \frac{dx^-}{4\pi}\langle P^\prime,S^\prime|
\bar{q}(-\frac{x^-}{2},{\bf 0}_\perp)
\gamma^+\gamma_5 q(\frac{x^-}{2},{\bf 0}_\perp)
|P,S\rangle 
e^{ix\bar{p}^+x^-}
\nonumber\\
& &\quad \quad
= \frac{1}{2\bar{p}^+}
\bar{u}(p^\prime,s^\prime)\left(\gamma^+\gamma_5  
\tilde{H}_q(x,\xi,t)
+ i\frac{\gamma_5\Delta^+}{2M}\tilde{E}(x,\xi,t)
\right)u(p,s)
\label{eq:gpd2}
\ee
with $\bar{p}^\mu = \frac{1}{2}\left( p^\mu
+p^{\prime \mu}\right)$ being the mean momentum
of the target,
$ \Delta^\mu = p^{\prime \mu}-p^\mu$ the four 
momentum transfer, and $t=\Delta^2$ the invariant
momentum transfer. The skewedness parameter 
$\xi = -\frac{\Delta^+}{2\bar{p}^+}$ quantifies 
the change in light-cone momentum.

GPDs allow for a unified description of a number of
hadronic properties \cite{xdj}; for example:
\begin{itemize}
\item[1.] In the forward
limit they reduce to conventional PDFs
\be
H_q(x,0,0)&=& q(x) \nonumber\\
\tilde{H}_q(x,0,0)&=& \Delta q(x).
\label{eq:forward}
\ee
\item[2.] When one integrates GPDs over $x$ they 
reduce to the usual form factors, e.g. the Dirac form
factors\footnote{Note that the dependence on the
longitudinal momentum transfer drops out due
to Lorentz invariance in Eq. (\ref{eq:form}).}
\be
\sum_q e_q \int dx H_q(x,\xi,t) &=& F_1(t) 
\nonumber\\
\sum_q e_q \int dx E_q(x,\xi,t) &=& F_2(t),
\label{eq:form}
\ee
\item[3.] Similar relations exist for the so called
`generalized form factors' 
\be
\sum_q e_q^2
\int \frac{dx}{x} H_q(x,0,t) &=& R_V(t) \nonumber\\
\sum_q e_q^2
\int \frac{dx}{x} E_q(x,0,t) &=& R_V(t) \nonumber\\
\sum_q e_q^2
\int \frac{dx}{x} \tilde{H}_q(x,0,t) &=& R_A(t) 
\nonumber\\
\sum_q e_q^2
\int \frac{dx}{x} \tilde{E}_q(x,0,t) &=& R_P(t) ,
\label{eq:gff}
\ee
which play a role in (real) wide angle Compton 
scattering \cite{wac}.
\item[4.] And integrals like
\be
{\cal H}_q(\xi,t)&\equiv&
\int_{-1}^1 dx H_q(x,\xi,t)\left(
\frac{1}{x+\xi-i\varepsilon}+
\frac{1}{x-\xi+i\varepsilon}\right)
\nonumber\\ & &
\label{eq:dvcs}
\ee
describe the deeply virtual Compton scattering
amplitude in the Bjorken scaling limit \cite{xdj}.
\end{itemize}
Even though the abovementioned experiments provide
only an indirect measurement of GPDs in the form of 
integrals (with the exception of the forward limit
of GPDs, which is measured in DIS), it is clear that 
GPDs parameterize many hadronic property at the same 
time. The fact that they are connected to so many
measurable quantities is not only the reason why
they are of such central importance for hadron
structure but at they same time provides a realistic
hope that a combination of data from different
classes of experiments together with enough
theoretical constraints will eventually suffice to
pin down these observables.

However, even though GPDs are such important 
observables, their physical interpretation is still 
rather obscure: since the initial and final states 
in Eqs. (\ref{eq:gpd}-\ref{eq:gpd2}) 
are not the same, GPDs in 
general do not have an immediate probabilistic 
interpretation as a density, which has made it
very difficult to develop a simple physical 
interpretation for these observables.

Eq. (\ref{eq:form}) already points towards a very
interesting connection between form factors and
GPDs in the sense that GPDs provide a decomposition
of the form factor with respect to the mean
longitudinal momentum fraction 
${x} = ({k_q^+ + {k_q^+}'})/{2\bar{P}^+}$
of the active quark, i.e. GPDs allow to determine 
how much quarks with a specific momentum fraction 
${x}$ contribute to the form factor and therefore
GPDs provide a (light-cone) momentum decomposition
of the form factor. Therefore, it should be clear 
that knowledge of GPDs should be able to 
discriminate between different mechanisms for 
form factors at large momentum transfer.

However, there is an even more important analogy 
between form factors and GPDs: the form-factor, i.e.
the non-forward matrix element of the current
operator, describes how the charge (i.e. the
forward matrix element of the same operator) is
distributed in position space.
\footnote{Strictly speaking this is of course only 
true nonrelativistically as well as in specific
frames (e.g. Breit frame), but this does not matter 
here since we use the connection 
$F(t)\longleftrightarrow \rho({\vec r})$ here only
as a motivation.}
GPDs are the off-forward matrix elements of the
operator
$\hat{O}_q(x,{\bf 0}_\perp)$ [Eq. (\ref{eq:0perp})].
The forward matrix elements of 
$\hat{O}_q(x,{\bf 0}_\perp)$ yield the usual PDFs.
By analogy with form factors one would therefore
expect that GPDs contain information about how
the usual PDFs (the forward matrix elements) are
distributed in position space \cite{ralston} (Table \ref{fig:rho}).
\begin{table}[h]
\tbl{Comparison between forward and off-forward
matrix elements of the current operator as well as
the operator that probes parton distribution 
functions. By analogy with form factors, one expects
that GPDs contain information about how PDFs
(the corresponding forward matrix elements) are
distributed in position space.}
{\begin{tabular}{@{}cccc@{}} \toprule
Operator & forward matrix & 
off-forward & position space \\
 & element & matrix element & interpretation \\
\colrule
$\bar{q}\gamma^+q$ & $Q$ 
 & $F(t)$ & $\rho({\vec r})$ \\[1.ex]
$\int\! \frac{dx^- e^{ixp^+x^-}\!\!\!\!\!}{4\pi}\,
\bar{q}\!\left(\!\frac{-x^-}{2}\!\right)\gamma^+
q\!\left(\!\frac{x^-}{2}\!\right)$
& $q(x)$ & $H_q(x,0,t)$ & $q(x,{\bf b_\perp})$ \\ \botrule
\end{tabular}}
\label{fig:rho}
\end{table}
The mere fact that some kind of connection between
transverse positions of partons and the Fourier
transform of GPDs with respect to the transverse
momentum transfer might exist is evident. 
However, what is less obvious
are important issues such as
`what exactly does one mean by the distribution of
partons in the transverse plane', `how is
that transverse distribution related to GPDs',
`what are the limitations and corrections (e.g.
relativistic corrections) to this position space
interpretation', `is there a strict probability
interpretation', and `what is the role of
polarization'.
The rest of the paper will be devoted to discussing 
this connection and and illustrating its 
consequences.
Before doing so, it is worthwhile to
discuss what kind of result one may anticipate.
First of all, since a measurement of a parton 
distribution corresponds to a measurement of the
momentum component of partons in the direction of the
target's momentum. Because of the Heisenberg 
uncertainty principle, one should therefore not 
expect to be able to measure the `longitudinal' 
position of partons. Measuring the 
transverse position simultaneously with the 
longitudinal momentum is not ruled out by the 
uncertainty principle, and therefore at the very
best what one might expect to able to determine
is the distribution of partons in the transverse 
plane. For this purpose, we will start out in the
following by defining what we mean by 
impact parameter dependent parton distributions
and then establish their connection to GPDs.

\section{Impact Parameter Dependent Parton 
Distribution}
\label{sec:impact}
In the case of nonrelativistic form factors, before
one can introduce the notion of a charge distribution
in position space, it is necessary to localize the
target, e.g. by working in the {\it center of mass}
frame. If one wants to talk about the distribution
of partons in position space then

In order to be able to define impact parameter
dependent PDFs one first needs to localize the
nucleon in the transverse direction. For this purpose
we introduce \footnote{Strictly speaking, we should work with
wave packets here in order to avoid states that
are normalized to $\delta$ functions.
This has been done in Ref. \cite{mb1} and will not
be repeated here because it makes the reasoning
much more lengthy and less transparent.}
\be
\left|p^+,{\bf R}_\perp= {\bf 0}_\perp,
\lambda\right\rangle
\equiv {\cal N}\int \frac{d^2{\bf p}_\perp}{(2\pi)^2} 
\left|p^+,{\bf p}_\perp, \lambda \right\rangle.
\label{eq:loc}
\ee
where $\left|p^+,{\bf p}_\perp, \lambda \right\rangle$
are light-cone helicity eigenstates
(see \ref{sec:galilei}\cite{soper}) and
${\cal N}$ is a normalization factor satisfying
$\left|{\cal N}\right|^2\int \frac{d^2{\bf p}_\perp}{(2\pi)^2}=1$.
This state is localized in the sense
that its {\sl transverse center of momentum}
${\bf R_\perp}$
is at the origin. For a state with total
momentum $p^+$ the transverse center of
momentum is defined as
\be
{\bf R_\perp} \equiv \frac{1}{p^+}
\int dx^- d^2{\bf x_\perp} \Theta^{++} {\bf x_\perp},
\ee
where $\Theta^{\mu \nu}$ is the energy momentum 
tensor (\ref{sec:galilei}). 
For practical purposes 
the intuitive parton representation for
${\bf R_\perp}$ is very useful
\be
{\bf R_\perp} =  \sum_i x_i {\bf r}_{\perp,i}.
\label{eq:Rperp}
\ee
The summation in Eq. 
(\ref{eq:Rperp}) is over all partons in the hadron 
and $x_i$ is the momentum fraction carried by the 
$i^{th}$ partons in the infinite momentum frame.
In summary, $\left| p^+, {\bf R_\perp}={\bf 0_\perp},
\lambda\right\rangle$ is a simultaneous eigenstate of
$\hat{p}^+$, ${\bf R_\perp}$ and $J_z$,
with eigenvalues $p^+$, ${\bf 0_\perp}$ and 
$\lambda$ respectively.

Working with this localized state is very similar to
working in the center of mass frame in 
nonrelativistic physics. 
The main difference being the weight factors, which 
are the mass fractions in the case of the 
nonrelativistic center of mass become momentum 
fractions in the transverse center of momentum. This 
is related to the properties of the Galilean 
subgroup of transverse boosts in the infinite 
momentum frame (\ref{sec:galilei}).

For such a transversely localized state, one can
define an {\bf impact parameter dependent PDF}, via
\footnote{
Note that for unpolarized parton distributions,
it is irrelevant whether
$\lambda=\uparrow$ or $\lambda=\downarrow$
in Eq. (\ref{eq:def1}).}
\be
q(x,{\bf b_\perp}) \equiv 
\left\langle p^+,{\bf R}_\perp= {\bf 0}_\perp,
\lambda\right|
\hat{O}_q(x,{\bf b_\perp})
\left|p^+,{\bf R}_\perp= {\bf 0}_\perp,
\lambda\right\rangle, 
\label{eq:def1}
\ee
where $\hat{O}_q(x,{\bf b_\perp})$ is obtained by a 
translation of Eq. (\ref{eq:0perp}) in the 
transverse plane, i.e.
\be
\hat{O}_q(x,{\bf b_\perp}) = 
\int \frac{dx^-}{4\pi}\bar{q}
\left(-\frac{x^-}{2},{\bf b_\perp} \right) \gamma^+ 
q\left(\frac{x^-}{2},{\bf b_\perp}\right) 
e^{ixp^+x^-}.
\label{eq:bperp}
\ee 
In gauges other than light cone gauge, a
straight line gauge string connecting the
points $(\frac{-x^-}{2}\!,{\bf b_\perp})$ and
$(\frac{x^-}{2}\!,{\bf b_\perp})$ 
needs to be inserted in Eqs. (\ref{eq:0perp}) and
(\ref{eq:bperp}).

In Section \ref{sec:prob} we will show that these
impact parameter dependent parton distributions
satisfy positivity constraints which justifies a
probabilistic interpretation. However, before that
we will establish the connection between impact
parameter dependent parton distributions and GPDs.

One may be tempted to expect a connection between 
impact parameter dependent PDFs, which are densities 
in transverse position space, and so called 
{\it unintegrated parton densities}
$f(x,{\bf k}_\perp)$,\cite{upd}
which are densities in transverse momentum space.
Even though there exist some inequalities, similar
to the Heisenberg inequality, relating the width in 
transverse position space to the width in transverse 
momentum space, no direct connection exists between 
the two, unless one makes specific assumptions about 
the functional form of the hadron wave function.
Strictly speaking, there just is no direct connection
between momentum densities and position space 
densities in a many particle system and therefore 
there is also no direct connection between 
$f(x,{\bf k}_\perp)$ and $q(x,{\bf b_\perp})$.

\subsection{Impact Parameter Dependent PDFs and GPDs}
In the following, we will focus on the case
of purely transverse momentum transfer $\Delta^+=0$ 
i.e. $\xi=0$. 
Furthermore, we will start out by considering
 GPDs without helicity flip where
only $H$ contributes in Eq. (\ref{eq:gpd}). 
Using
$\bar{u}(p^\prime)\gamma^+u(p) = 2p^+$ when 
$p^{\prime +}=p^+$ one thus finds
\be
& &\!\!\!\!\!\!\!\!
\int \frac{dx^-}{4\pi} \langle P^\prime,\lambda|
\bar{q}(-\frac{x^-}{2},{\bf 0}_\perp)
\gamma^+ q(\frac{x^-}{2}\!,{\bf 0}_\perp)|P,\lambda \rangle 
e^{ixp^+x^-} =  H_q(x,t),
\label{eq:Hx0t}
\ee
where we introduced the
shorthand notation $H_q(x,t)\equiv H_q(x,0,t)$.

In order to illustrate the physics of $H_q(x,t)$,
we take the definition of impact parameter
dependent PDFs and expand the definition
of the nucleon state in the center of momentum
frame  
$\left| p^+,{\bf R_\perp}={\bf 0_\perp},
\lambda\right\rangle$
in a plane wave basis  
\be 
q(x,{\bf b_\perp}) &\equiv& 
\left\langle p^+,{\bf R_\perp}={\bf 0_\perp},
\lambda \right| \hat{O}_q(x,{\bf b_\perp})
\left| p^+,{\bf R_\perp}={\bf 0_\perp},
\lambda\right\rangle 
\label{eq:step1}
\\
&=&
\left|{\cal N}\right|^2
\int\!\!\! \frac{d^2 {\bf p}_\perp }{(2\pi)^2}
\!\int \!\!\!\frac{d^2 {\bf p}_\perp^\prime}{(2\pi)^2}
\left\langle p^+\!,
{\bf p}_\perp^\prime,\lambda \right|
 \hat{O}_q(x,{\bf b_\perp}) \left|p^+\!,
{\bf p}_\perp,\lambda \right\rangle
\nonumber\\
&=&\left|{\cal N}\right|^2 
\int \!\!\!\frac{d^2 {\bf p}_\perp}{(2\pi)^2}\! \int \!\!\!\frac{d^2 
{\bf p}_\perp^\prime}{(2\pi)^2}
\left\langle p^+\!,{\bf p}_\perp^\prime,\lambda 
\right|
 \hat{O}_q(x,{\bf 0_\perp}) \left|p^+\!,
{\bf p}_\perp \lambda \right\rangle 
 e^{i{\bf b_\perp}
\cdot ({\bf p}_\perp-{\bf p}_\perp^\prime)}
e^{ixp^+x^-}
\nonumber
,
\ee
where the phase factor in the last step is due to
the transverse translation, i.e.
$
\left\langle p^+\!,{\bf p}_\perp^\prime \right|
\hat{O}_q(x,{\bf b_\perp})\left|p^+\!,
{\bf p}_\perp \right\rangle
= 
\left\langle p^+\!,{\bf p}_\perp^\prime \right|
\hat{O}_q(x,{\bf 0}_\perp)\left|p^+\!,
{\bf p}_\perp \right\rangle
 e^{i{\bf b_\perp}
\cdot ({\bf p}_\perp-{\bf p}_\perp^\prime)}$.
The kind of matrix elements that appear in
Eq. (\ref{eq:step1}) are identical to matrix
elements that appear in the definition of 
$H_q(x,0,t)$. Combining Eqs. (\ref{eq:step1}) and
(\ref{eq:Hx0t}) one thus finds 
\be 
q(x,{\bf b_\perp}) 
&=& \left|{\cal N}\right|^2
 \int \frac{d^2{\bf p}_\perp}{(2\pi)^2}  \int 
\frac{d^2{\bf p}_\perp^\prime}{(2\pi)^2}
H_q(x,-\left({\bf p}_\perp-{\bf p}_\perp^\prime
\right)^2) 
e^{i{\bf b_\perp}
\cdot ({\bf p}_\perp-{\bf p}_\perp^\prime)}
\nonumber\\
&=&
 \int \frac{d^2{\bf \Delta}_\perp}{(2\pi)^2}  
H_q(x,-{\bf \Delta}_\perp^2) e^{-i{\bf b_\perp} \cdot
{\bf \Delta}_\perp},
\label{eq:result1}
\ee
where we switched in the last step to total and
relative transverse momentum, i.e.\\
$\int d^2{\bf p}_\perp \int d^2{\bf p}_\perp^\prime
= \int d^2{\bf \Delta}_\perp \int 
d^2{\bf \bar{p}}_\perp$ 
, with
${\bf \bar{p}}_\perp = \frac{1}{2}
\left({\bf p}_\perp + {\bf p}_\perp^\prime \right)$
and
${\bf \Delta}_\perp = {\bf p}_\perp^\prime - 
{\bf p}_\perp $,
and used the fact that $H_q$ did not depend on
${\bf \bar{p}}_\perp$.\footnote{Note that the last 
property, i.e. the fact that 
$H_q(x,0,-{\bf \Delta}_\perp^2)$ does not depend on 
${\bf \bar{p}}_\perp$ (which reflects the Galilei 
invariance under $\perp$ boosts in the IMF), is one 
of the crucial ingredients of our derivation!}

This proves that the impact parameter dependent PDF
defined in Eq. (\ref{eq:def1}) is the Fourier 
transform of $H_q(x,-{\bf \Delta}_\perp^2)$, i.e.
if one knows $H_q(x,-{\bf \Delta}_\perp^2)$ one can
determine the distribution of partons simultaneously
as a function of the light-cone momentum fraction 
$x$ and the distance ${\bf b_\perp}$ from the 
transverse center of momentum.\footnote{If one
includes a factor $(1-x)$ in the exponent of the
Fourier transform in Eq. (\ref{eq:result1}), one
obtains the distribution not as a function of
the distance of the active quark from the CM of
the whole hadron but as a function of the distance
to the CM of the spectators.}
This very important result was already obtained in 
Ref. \cite{mb1}, where it was shown that working 
with wave 
packets in $\perp$ momentum space, i.e. replacing
$\int d^2{\bf p}_\perp$ by $\int d^2{\bf p}_\perp
\psi({\bf p}_\perp)$, where $\psi({\bf p}_\perp)$
is a slowly varying function of ${\bf p}_\perp$
that is taken to be a constant towards the end of the
calculation, does not change the final result.
\footnote{This result was also confirmed in Ref.
\cite{diehl2}, where Gaussian wave packets were 
used.}
It only leads to much more complicated intermediate
expressions.
Here we followed a much more simplified 
derivation\cite{mb2} using non-normalizable states
in order to better 
illustrate the physics of the result.

As a side remark, it should be emphasized that the
interpretation of the Fourier transform of form
factors as charge densities is, in the rest frame,
spoiled by relativistic normalization factors as
well as the effect of Lorentz contraction. It is
also well known that these relativistic corrections 
are not present in special  frames, such as for 
example the Breit frame. Our derivation above shows 
that such corrections not present in the infinite 
momentum frame either, i.e. the identification of 
the Fourier transform of $H_q(x,0,-{\bf \Delta}^2)$ 
w.r.t. ${\bf \Delta}$ as a distribution of partons 
in transverse position space is {\sl not} limited by 
relativistic effects. As a corollary, since 
$\sum_q e_q \int dx 
H_q(x,0,-{\bf \Delta}^2)= F_1(-{\bf \Delta}^2)$, 
this means that the Fourier transform of 
$F_1(-{\bf \Delta}^2)$ can be interpreted as the 
charge distribution in the infinite momentum frame 
as a function of the transverse distance from the 
(transverse) center of momentum. This interpretation,
together with the observation that this 
interpretation is not spoiled by relativistic 
corrections, are both surprisingly little known results.

\subsection{Density Interpretation for Impact 
Parameter Dependent Parton Distributions}
\label{sec:prob}
In this section, we will show that 
$q(x,{\bf b_\perp})$ satisfies positivity 
constraints, which allows one to associate a 
probabilistic interpretation. For this purpose we 
note that only the projection on the `good' quark 
field component 
$\psi_{(+)} \equiv \frac{1}{2}\gamma^-\gamma^+ \psi$
contributes in
twist-2 matrix elements, i.e. 
$\bar{\psi}^\prime \gamma^+\psi =
\bar{\psi}^\prime_{(+)} \gamma^+\psi_{(+)}$.

When one quantizes fermions on the light-cone (or in 
the infinite momentum frame) only this `good' component
is dynamical \footnote{The `bad' component
$\psi_{(-)} \equiv \frac{1}{2}\gamma^+\gamma^- \psi$
satisfies a constraint equation.}
and an expansion in terms of canonical 
raising and lowering operators reads 
\cite{mb:adv}
\bea
\psi_{(+)}(x^-\!,{\bf x}_\perp)
= \int_0^\infty \!\!\!\frac{dk^+}{\sqrt{4\pi k^+}} 
\int \!\!\frac{d^2{\bf k}_\perp}{2\pi}
&\sum_s& \left[
u_{(+)}(k,s) b_s(k^+\!,{\bf k}_\perp)
e^{-ikx} \right.\nonumber\\
& &\left.+ v_{(+)}(k,s)
d_s^\dagger(k^+\!,{\bf k}_\perp) e^{ikx}\right],
\ee
where $b$ and $d$ satisfy the usual (canonical)
equal light-cone
time $x^+$ anti-commutation relations, e.g.
\be
\left\{ b_r(k^+\!,{\bf k}_\perp),
 b^\dagger_s(q^+\!,{\bf q}_\perp)\right\} =
\delta(k^+\!-q^+) 
\delta({\bf k}_\perp\!-{\bf q}_\perp)
\delta_{rs}
\ee
and the normalization of the spinors is such that
\be
\bar{u}_{(+)}(p,r)\gamma^+u_{(+)}(p,s)=
2p^+\delta_{rs}.
\ee
An explicit representation for these spinors can be 
found in Ref. \cite{soper}. We now insert these 
expansions into our definition of 
$q(x,{\bf b_\perp})$. Using 
$
\bar{u}_{(+)}(p^\prime,r)\gamma^+ u_{(+)}(p,s) = 
2p^+\delta_{rs},
$
when $p^+={p^+}^\prime$, one finds for $x>0$
\bea
q(x,{\bf b_\perp}) &=&
{\cal N}^\prime
\sum_s\! \int \!\!\frac{d^2{\bf k}_\perp}{2\pi}
\!\!\int \!\!\frac{d^2{\bf k}_\perp^\prime}{2\pi}
e^{-i{\bf b_\perp}\cdot({\bf k}_\perp^\prime -
{\bf k}_\perp)}\nonumber\\
& &\left\langle p^+,
{\bf R_\perp}={\bf 0_\perp},\lambda
\right|
b_s^\dagger(xp^+\!,{\bf k}_\perp^\prime)
b_s(xp^+\!,{\bf k}_\perp)
\left|p^+,{\bf R_\perp}={\bf 0_\perp},\lambda
\right\rangle 
.
\label{eq:dens1}
\ee
Note that we made use of the fact that the states 
that appear in the `initial' and `final' states of 
Eq. (\ref{eq:dens1}) have the same $p^+$ and 
therefore both the $k^+$ of the lowering and the 
raising operator must be the same (i.e. both are 
equal to $xp^+$), which allows one to replace in the 
matrix element
\be
\int \!\!\frac{dx^-}{4\pi}
\bar{q}(0^-\!,{\bf b_\perp})\gamma^+
q(x^-\!,{\bf b_\perp})e^{ix^-p^+x} 
\quad\longrightarrow \quad {\cal N}^\prime\,
\tilde{b}^\dagger (xp^+,{\bf b_\perp})
\tilde{b}(xp^+,{\bf b_\perp}),
\ee
where ${\cal N}^\prime$ is a constant, which is, in 
the infinite volume proportional to $\delta(0)$. For 
$x<0$ one finds a similar expression with
$b_s^\dagger b_s$ replaced by  $-d_sd_s^\dagger$.

In order to simplify these expressions further, we 
introduce a `hybrid' representation, with field 
operators that are labeled by $\perp$ position space 
and longitudinal momentum space variables, i.e. we 
introduce
\be
\tilde{b}(k^+,{\bf x}_\perp) \equiv
\int \frac{d^2{\bf k}_\perp}{2\pi} 
b(k^+,{\bf k}_\perp) e^{i{\bf k}_\perp \cdot 
{\bf x}_\perp}
\ee
and likewise for $\tilde{d}^\dagger$. Inserting 
$\tilde{b}$ into Eq. (\ref{eq:dens1}) one easily 
finds
\be
q(x,{\bf b_\perp}) &\sim&
\sum_s
\left\langle p^+,{\bf R_\perp}={\bf 0_\perp},\lambda
\right|
\tilde{b}_s^\dagger(xp^+,{\bf b_\perp})
\tilde{b}_s(xp^+,{\bf b_\perp})
\left|p^+,{\bf R_\perp}={\bf 0_\perp},\lambda
\right\rangle 
\nonumber\\
&=&
\sum_s \left| \tilde{b}_s(xp^+,{\bf b_\perp})
\left|p^+,{\bf R_\perp}={\bf 0_\perp},\lambda
\right\rangle \right|^2 
\geq 0,
\label{eq:dens2}
\ee
for $x<0$. Similarly, one finds for $x<0$
\be
q(x,{\bf b_\perp})&\sim& -
\sum_s \left| \tilde{d}^\dagger_s(xp^+,{\bf b_\perp})
\left|p^+,{\bf R_\perp}={\bf 0_\perp},\lambda
\right\rangle \right|^2 
\leq 0.
\label{eq:dens3}
\ee
These results were first shown in Ref. \cite{mb2} and
confirmed in Ref. \cite{diehl2}.

Because of these positivity (negativity) constraints,
it is legitimate to associate the physical
interpretation of a probability density distribution
with the impact parameter dependent parton
distribution $q(x,{\bf b_\perp})$. The fact that the 
impact parameter dependent parton distributions 
have a density interpretation is very important 
because it shows that they have a physical 
significance beyond being the Fourier transforms of
GPDs!

\subsection{Modeling Impact Parameter Dependent 
Parton Distributions}
GPDs cannot be uniquely determined from Compton
scattering experiments alone, where they appear
in the integrand of convolution integrals. Until
sufficient constraints from theory and/or other
experiments exist, that allow one to pin down
GPDs uniquely, one may still attempt to extract
GPDs from (virtual) Compton scattering data by
making a physically motivated ansatz for the
GPDs, where the free parameters in the ansatz are
determined by fitting to the data. Since such
a procedure is clearly model dependent, it is
desirable to use as many general theoretical
constraints as possible in order to minimize
the dependence on specific model features.
The results from this paper can be used to
provide some general theoretical constraints 
as we will discuss in the following.

First of all, in order to be consistent with the 
above positivity constraints [(\ref{eq:dens2}) and 
(\ref{eq:dens3})], all realistic Ans\"atze for 
$H_q(x,0,t)$ should satisfy 
\be
\int d^2{\bf \Delta}_\perp e^{-i{\bf b_\perp}\cdot {\bf \Delta}_\perp}
H_q(x,0,-{\bf \Delta}_\perp^2) &\geq& 0 \quad \quad\mbox{for} \quad
x>0 \nonumber\\
\int d^2{\bf \Delta}_\perp e^{-i{\bf b_\perp}\cdot {\bf \Delta}_\perp}
H_q(x,0,-{\bf \Delta}_\perp^2) &\leq& 0 \quad \quad\mbox{for} \quad
x<0 
\ee
for any value of ${\bf b_\perp}$ and any value of 
$x$.
Of course, many Ans\"atze, especially those with a 
Gaussian $t$ dependence, are already consistent with 
these new constraints. However, additional 
constraints arise due to the fact that GPDs can be 
linked to impact parameter dependent PDFs.

First of all, the mere fact that the impact parameter
dependent PDFs are measured with respect to the
transverse center of momentum ${\bf R}_\perp= 
\sum_i x_i {\bf r}_{\perp,i}$ implies that 
$H_q(x,0,t)$ should become $t$-independent as 
$x\rightarrow 1$.  The reason for this behavior is
very simple: the transverse center of momentum of a
hadron is defined as the weighted average of 
transverse positions of all partons in the hadron, 
where the weight factors are the momentum fractions 
carried by each parton. Since $x$ in $H_q(x,0,t)$ is 
the momentum fraction carried by the active quark,
this active quark carries all the momentum of the 
hadron when one takes $x\rightarrow 1$ and therefore 
the contribution from all partons other than the 
active quark in Eq. (\ref{eq:Rperp}) becomes 
negligible. Hence the active quark is always very 
close to the transverse center of momentum when 
$x\rightarrow 1$ and the impact parameter dependent 
PDF $q(x,{\bf b_\perp})$ has a very peaked 
`transverse profile' in this limit. More precisely, 
the transverse width of the impact parameter 
dependent PDF
\be
\langle R_\perp^2(x)\rangle \equiv 
\frac{\int d^2{\bf b_\perp}
{\bf b}_\perp^2 q(x,{\bf b_\perp})}
{\int d^2{\bf b_\perp}
q(x,{\bf b_\perp})}
\ee
should vanish as $x\rightarrow 1$.
\footnote{This is very similar to $B$-mesons, where 
the $b$ quark does not recoil very much and therefore
has a very localized distribution in position space.}
Thus $H_q(x,0,-{\bf \Delta}_\perp^2)$ (being the
Fourier transform of $q(x,{\bf b_\perp})$) should 
become independent of
${\bf \Delta}_\perp^2$ as $x\rightarrow 1$.
We should emphasize that this behavior is 
consistent with the 'Feynman picture' for hadron 
form factors near $x=1$ \cite{feynman}. In this 
picture, configurations where one parton carries 
$x\approx 1$ and all others are `wee' can be easily 
scattered elastically (by striking the parton with 
$x\approx 1$) since the wee partons do not have a 
direction and thus can be easily `turned
around'. 
Therefore, in the Feynman picture, contributions to 
the form factor where the active quark carries a 
larger fraction of the hadron's total momentum tend 
to have a weaker $t$ dependence.

The constraint that $H_q(x,0,t)$ should become
$t$ independent as $x\rightarrow 1$ rules out
factorized Ans\"atze of the form
\be
H_q(x,0,t) = q(x) \cdot F(t),
\ee
since they do not become $t$ independent near
$x=1$.

The position space interpretation of GPDs also
allows one to put constraints on the
behavior near $x=0$. Space time descriptions
of hadron structure \cite{gribov} suggest that
the transverse size of hadrons should grow
like $\alpha \ln \frac{1}{x}$ near $x=0$, where
$\alpha$ is some constant. In fact, a more
rapid growth (e.g. power law) of the transverse
size for $x\rightarrow 0$ would also lead to
power law growth of total cross sections 
and would therefore lead to violations of unitarity 
bounds.

Because of the connection between the transverse
size and the slope of $H_q(x,0,t)$, this result 
implies for example that an ansatz like
\be
H_q(x,0,-{\bf \Delta}_\perp^2) = 
q(x) e^{-a{\bf \Delta}_\perp^2
\frac{1-x}{x}},
\label{eq:gauss}
\ee
is inconsistent with abovementioned constraints
about the transverse size of hadrons near $x=0$. 
\footnote{Of course, for intermediate and large 
values of $x$, this ansatz may nevertheless be a 
useful approximation to the actual behavior and this
ansatz also satisfies our above constraint that 
$H_q(x,0,t)$ should become $t$ independent near 
$x=1$.}

An improved Gaussian ansatz for 
$H_q(x,0,-{\bf \Delta}_\perp^2)$, which is consistent 
with both the $x\rightarrow 1$ and $x\rightarrow 0$ 
constraints, is given by
\be
H_q(x,0,-{\bf \Delta}_\perp^2) = q(x)
e^{-a{\bf \Delta}_\perp^2 \ln \frac{1}{x}}
\label{eq:log1}
\ee
or
\be
H_q(x,0,-{\bf \Delta}_\perp^2) = q(x)
e^{-a{\bf \Delta}_\perp^2(1-x)\ln \frac{1}{x}} .
\label{eq:log2}
\ee
The latter ansatz yields at the same time an
asymptotic form factor that is consistent with
quark counting: if the end-point behavior 
of the PDF is taken such that
$q(x) \sim (1-x)^{2n_s-1}$ near $x=1$ then 
the large $Q^2$ power law behavior of 
the form factor is
$F(Q^2) \sim \left(\frac{1}{Q^2}\right)^{n_s}$.

Although the above discussion may appear purely
academic at the present stage, where very little
information on GPDs exists, it may still have some 
phenomenological significance because in Compton 
scattering the contributions from the small $x$ 
region are enhanced due to the factor $1/x$ 
(\ref{eq:gff}). The ansatz in Eq. (\ref{eq:gauss}) 
leads to a much more rapid suppression of 
contributions from the small $x$ region at larger 
values of $t$ than a logarithmic dependence on $x$ 
in the exponent [Eqs. (\ref{eq:log1}) and 
(\ref{eq:log2})].

\begin{figure}
\unitlength1.cm
\begin{picture}(10,8)(4.5,20.5)
\includegraphics{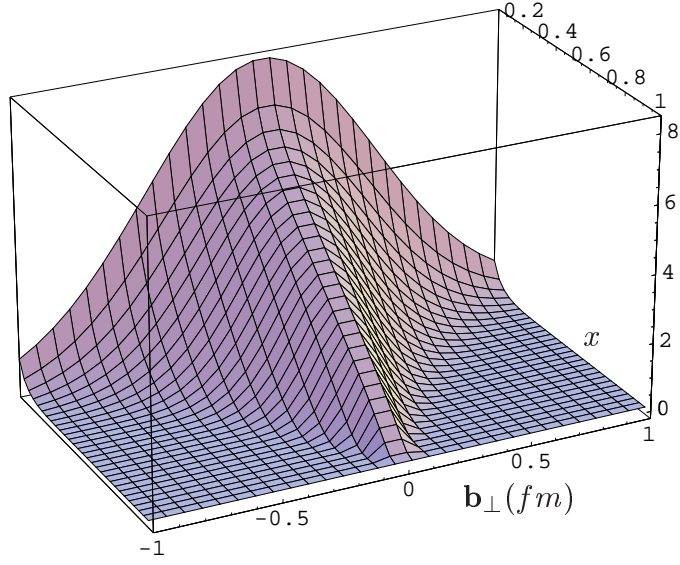}
\end{picture}
\caption{Impact parameter dependent parton distribution
$u(x,{\bf b_\perp})$ for the simple model 
(\ref{eq:model}).}
\label{fig:fgpd}
\end{figure}

In order to gain some intuitive understanding
about the behavior of impact parameter dependent
parton distributions, we transform Eq. (\ref{eq:log2})
to impact parameter space, where we find 
\be
q(x,{\bf b_\perp}) = q(x)\frac{1}{4\pi a (1-x)\ln \frac{1}{x}}
e^{-\frac{ {\bf b}_\perp^2 }{4a(1-x)\ln \frac{1}{x}}} .
\label{eq:model}
\ee
The result, which is depicted in Fig. \ref{fig:fgpd},
exhibits the general features that we described above.
First, there is a general decrease in magnitude with increasing
$x$, which arises mostly due to the decrease of $q(x)$.
Furthermore, one can also a notice a decreasing 
transverse
width of the distribution $q(x,{\bf b_\perp})$ as 
$x$ increases, which is in accord with our general
prediction. For $x\rightarrow 1$ the transverse width
goes to zero.

\section{Nucleon Helicity Flip Distributions}
\label{sec:flip}
\begin{figure}
\unitlength1.cm
\begin{picture}(14,8.2)(-1.5,7.8)
\includegraphics{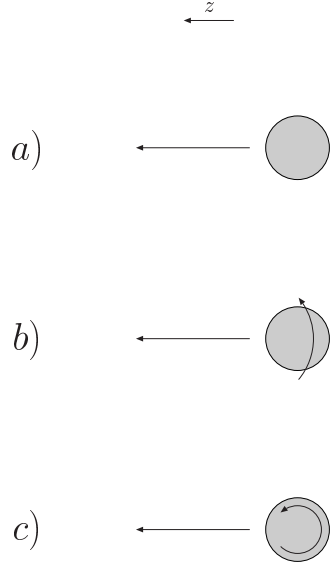}
\end{picture}
\caption{
Comparison of a non-rotating sphere that moves in 
the
$z$ direction 
with a sphere that spins at the same time
around the $z$ axis
and a sphere that spins around the $x$ axis
When the sphere spins around the $x$ axis, the
rotation changes the distribution
of momenta in the $z$ direction (adds/subtracts to
velocity for $y>0$ and $y<0$ respectively).
For the nucleon the resulting modification of the
(unpolarized) momentum distribution is 
described by $E(x,0,t)$.}
\label{fig:fspin}
\end{figure}

So far we have considered only two out of eight
GPDs \cite{flip}. In order to investigate the 
interpretation of the other six GPDs, it is necessary
to consider amplitudes where either the nucleon
or quark helicity (or both) flip, which makes it much
more difficult to develop a probabilistic 
interpretation. Therefore, we focus here on 
$E(x,0,t)$,
which appears in amplitudes where the nucleon
helicity flips and the quark helicity does not
flip 
\bea
 \label{eq:flip}
\int \frac{dx^-}{4\pi} e^{ip^+x^- x}
\left\langle P+\Delta, \uparrow
\left| \bar{q}\left(0\right)\gamma^+ 
q\left({x^-}\right)
\right| P,\downarrow\right\rangle 
&=& -\frac{\Delta_x-i\Delta_y}{2M}E(x,0,-{\bf \Delta}_\perp^2) .
\eea
For a probabilistic interpretation, it is
necessary to consider amplitudes where the initial
and final states are the same. Since $E(x,0,t)$
does not contribute when the initial and final
state have the same helicity, we will now consider
a state that is a superposition of 
(transversely localized) nucleon states with
opposite helicities. For brevity, we will denote
this state as $\left|X\right\rangle$, i.e.
\be
\left|X\right\rangle \equiv
\frac{1}{\sqrt{2}}\left(\left|p^+,{\bf R_\perp}={\bf 0_\perp},\uparrow
\right\rangle +
\left|p^+,{\bf R_\perp}={\bf 0_\perp},\downarrow
\right\rangle\right) .
\ee
It is tempting to interpret this state as a state
that is polarized in the $x$ direction and if we 
were dealing with a nonrelativistic system this
would certainly be the case. However, for 
relativistic systems, spins are not invariant under 
boosts and one has to be careful here with such
an interpretation. Nevertheless, let us proceed here
and study the (unpolarized)
impact parameter dependent PDF in
this state
\bea
q_X(x,{\bf b_\perp})
&\equiv&
\left\langle X
\right|
\hat{O}_q(x,{\bf b_\perp})
\left| X \right\rangle \label{eq:fsx}\\
&=& \int \frac{d^2{\bf \Delta_\perp}}{(2\pi)^2}
e^{-i{\bf \Delta_\perp}\cdot {\bf b_\perp}}
\left[
 H_q(x,0,-{\bf \Delta}_\perp^2) + i 
\frac{\Delta_y}{2M} E_q(x,0,-{\bf \Delta}_\perp^2)
\right]
\nonumber\\
&=& q(x,{\bf b_\perp})
-\frac{1}{2M} \frac{\partial}{\partial b_y}
{\cal E}_q(x,{\bf b_\perp})
 ,
\label{eq:qX}
\eea
where we denoted the Fourier transform of
$E(x,0,-{\bf \Delta}_\perp^2)$ by
${\cal E}(x,{\bf b_\perp})$, i.e.
\be
{\cal E}_q(x,{\bf b_\perp}) \equiv
\int \frac{d^2{\bf \Delta_\perp}}{(2\pi)^2}
e^{-i{\bf \Delta_\perp}\cdot {\bf b_\perp}}
E_q(x,0,-{\bf \Delta}_\perp^2).
\ee
The physical significance of ${\cal E}(x,{\bf b_\perp})$ is as
follows. First, integration over both
$x$ and ${\bf b_\perp}$ yields the 
contribution from quark flavor $q$ to the anomalous
magnetic moment
\be
\int_{-1}^1 dx \int d^2 {\bf b_\perp}
{\cal E}_q(x,{\bf b_\perp}) = F_{2,q}(0) = 
\kappa_q .
\ee
Secondly, Eq. (\ref{eq:fsx}) describes the 
transverse distortion
of the unpolarized impact parameter dependent PDF
if the state is not polarized in the $\pm z$
direction but rather in a transverse direction. 
The physics of this result can be
illustrated by some classical example 
(Fig. \ref{fig:fspin})
Consider a sphere that moves in the $z$ direction.
The analog of PDFs in this example are the
distributions of momenta in the $z$ direction
from particles on the sphere as seen by an
observer at rest. The distribution of these momenta
in the transverse plane is the same regardless
whether the sphere is nonrotating or spinning
around the $z$ axis (in either direction) because
spinning around the $z$ axis does not modify
the momenta in the $z$ direction.\footnote{One
may also take this example as a simple illustration
as to why the (unpolarized)
impact parameter dependent PDFs are the same for
nucleons with $\lambda=\uparrow$ and
$\lambda=\downarrow$.} However, if the sphere
spins around the $x$-axis while still moving along
the $z$-axis, the longitudinal
momentum distribution 
of particles on the sphere as
seen by an observer at rest changes dependent
on the transverse position of the observer since
on one side of the $z$ axis the rotational
motion adds to the translatory motion and on the 
other side it subtracts.

Of course, the nucleon is not just a rotating sphere
but this simple example illustrate the physics of why
the parton distribution in the transverse plane gets
distorted and is no longer invariant under rotations
around the $z$-axis when the nucleon is not 
longitudinally
polarized. And in the nucleon it is the Fourier
transform of $E(x,0,-{\bf \Delta}_\perp^2)$, which
describes this distortion.

The second moment of $q_X(x,{\bf b_\perp})$, i.e.
$\int_{-1}^1 dx x q_X(x,{\bf b_\perp})$, is of 
particular interest since it describes how
the momentum is distributed in the transverse plane
\be
\int_{-1}^1 dx x q_X(x,{\bf b_\perp})
= \int_{-1}^1 dx x q(x,{\bf b_\perp})
-
\frac{\partial}{\partial b_x}
\int_{-1}^1 dx\, x {\cal E}_q(x,{\bf b_\perp}),
\ee
i.e. the second moment of ${\cal E}_q$ describes 
how the distribution of momentum carried by 
quarks of flavor $q$ in the transverse plane
changes when the nucleon is not polarized in the
$z$ direction. 
In particular, the contribution from flavor $q$ to
the transverse center of momentum in the 
``transversely polarized'' state 
$\left|X\right\rangle$ is shifted away from the
origin by
\bea
\int_{-1}^1 dx\int d^2{\bf b_\perp}\,
b_y x\, q_X(x,{\bf b_\perp}) 
&=&
-\int_{-1}^1 dx \int d^2{\bf b_\perp}\,
b_y x \frac{1}{2M}
\frac{\partial}{\partial b_y}
{\cal E}_q(x,{\bf b_\perp})\nonumber\\
&=& \int_{-1}^1 dx \int d^2{\bf b_\perp}\,
x \frac{1}{2M}{\cal E}_q(x,{\bf b_\perp})
\nonumber\\
&=& \frac{1}{2M} \int_{-1}^1 dx\, x E_q(x,0,0) .
\eea
Therefore, the
second moment of $E_q(x,0,0)$ describes, in units
of $\frac{1}{2M}$, by how much the transverse center 
of momentum of quarks of flavor $q$ is shifted
away from the origin in the `transversely 
polarized' state $\left|X \right\rangle$.
Of course, when summed over all partons (all quarks 
and gluons) the transverse center of momentum of the
state $\left|X \right\rangle$ is still at the
origin since it is a superposition of eigenstates
of ${\bf R_\perp}$ with eigenvalue ${\bf 0_\perp}$,
i.e.
\be
\sum_{i\in q,g} \int_{-1}^1 dx\, x E_i(x,0,0)=0,
\ee
which is nothing but the statement that the total
(i.e. all flavors plus glue)
anomalous gravito-magnetic moment has to vanish
\cite{anom}.

If the center of momentum for quarks of flavor
$q$ is shifted away from the origin in the $y$
direction then those quarks carry orbital
angular momentum in the $x$ direction. Starting
from this observation, one can obtain a more
physical understanding of why the $2^{nd}$ moment
of $E$ appears in Ji's angular momentum sum rule
\cite{mb2}.

Another important consequence can be derived from
the fact that $q_X(x,{\bf b_\perp})$, i.e. the 
unpolarized impact parameter dependent quark 
distribution in the state $\left| X \right\rangle$
has a probabilistic interpretation as well
\bea
q_X(x,{\bf b_\perp}) &\geq& 0
\quad \quad \quad \mbox{for}\quad x>0
\nonumber\\
q_X(x,{\bf b_\perp}) &\leq& 0
\quad \quad \quad \mbox{for}\quad x<0 .
\eea
This implies the inequality
\be
q(x,{\bf b_\perp})
\geq \frac{1}{2M} \left|\frac{\partial}{\partial b_i}
{\cal E}_q(x,{\bf b_\perp}) \right| .
\label{eq:ineqE}
\ee
The physics of this result is that although
the probability distribution of partons in the
transverse plane gets distorted when the nucleon
is polarized in a transverse direction, the
resulting
probability density needs to remain positive.
Since the distortion is described by the gradient of
${\cal E}(x,{\bf b_\perp})$, 
this sets some upper bound on the magnitude of
${\bf \nabla_\perp}{\cal E}$.
By considering more general helicity combinations
one can derive inequalities connecting
chirally even and odd parton distributions
in impact parameter representation \cite{pub}, but
we will not consider those here since chirally
odd GPDs are even harder to measure than chirally 
even ones.

\begin{figure}
\unitlength1.cm
\begin{picture}(10,16.5)(2.3,3.5)
\includegraphics{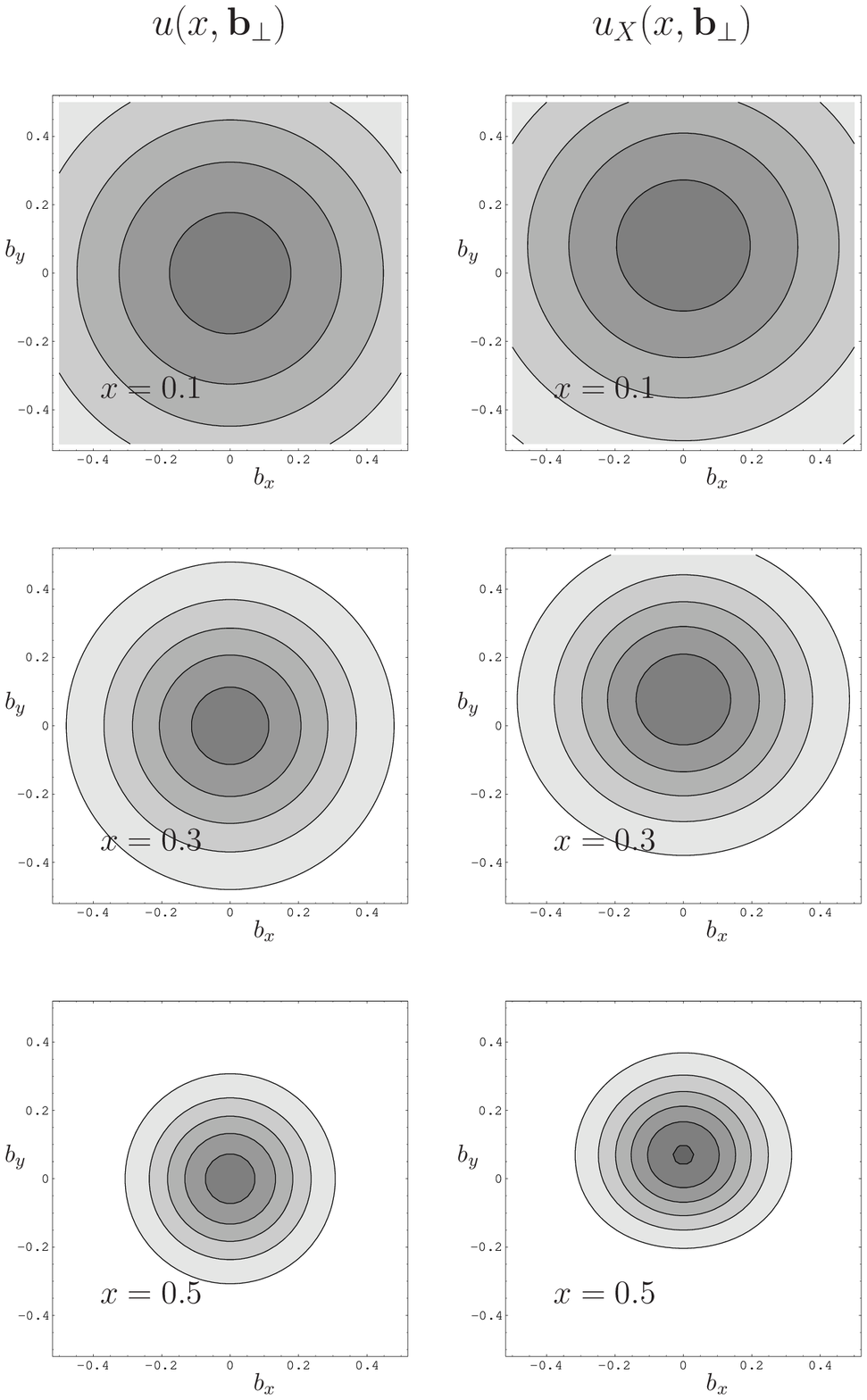}
\end{picture}
\caption{$u$ quark distribution in the transverse
plane for $x=0.1$, $0.3$, and $0.5$ (\ref{eq:model}).
Left column: $u(x,{\bf b_\perp})$, i.e. the
$u$ quark distribution for unpolarized 
protons; right column: $u_X(x,{\bf b_\perp})$, i.e.
the unpolarized $u$ quark distribution for
`transversely polarized'
protons $\left|X\right\rangle = 
\left|\uparrow\right\rangle +
\left|\downarrow\right\rangle$.
The distributions are normalized to the central 
(undistorted) value $u(x,{\bf 0_\perp})$.}
\label{fig:panelu}
\end{figure}
\begin{figure}
\unitlength1.cm
\begin{picture}(10,16.5)(2.3,3.5)
\includegraphics{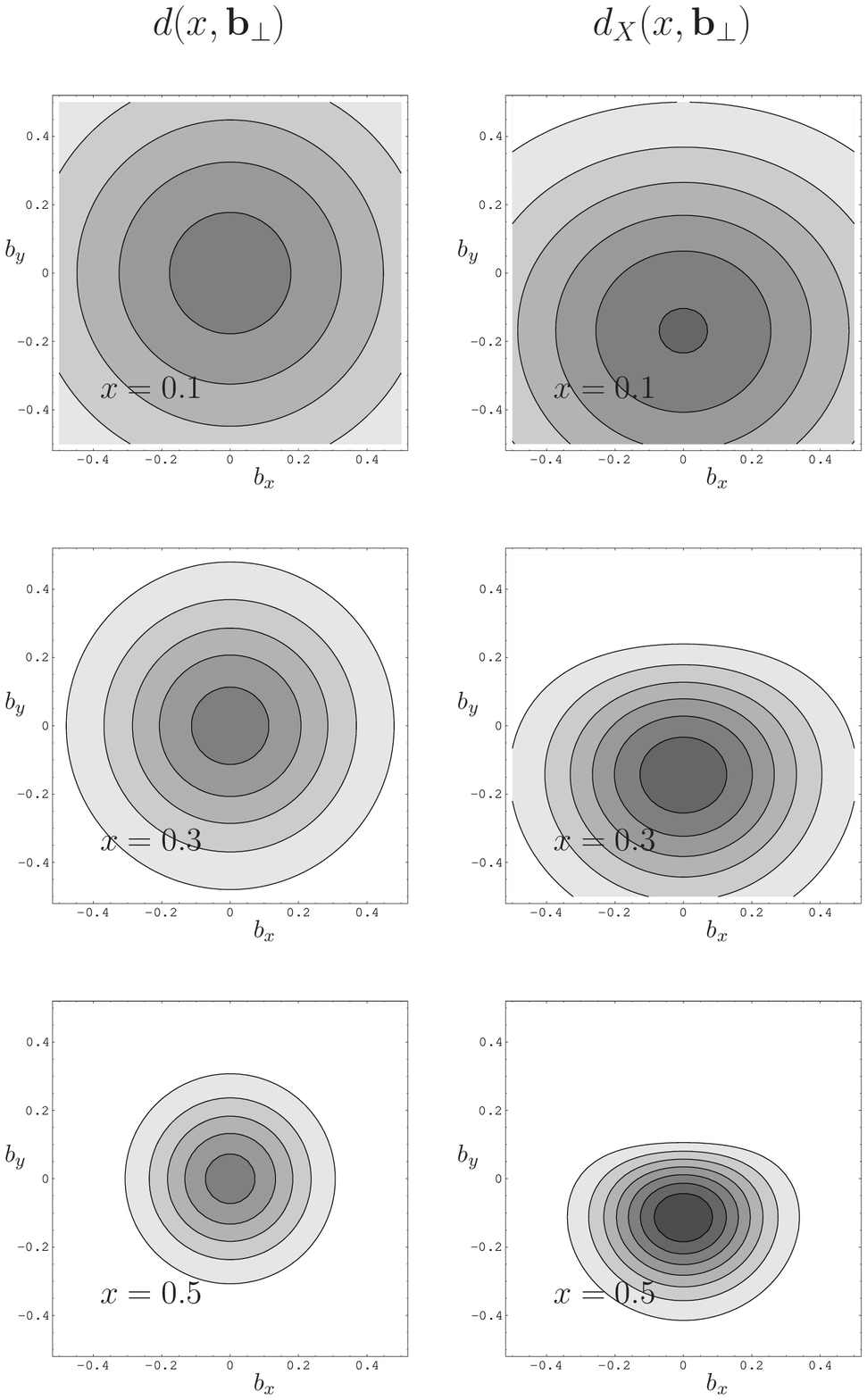}
\end{picture}
\caption{Same as Fig. \ref{fig:panelu}, but for
$d$ quarks. }
\label{fig:paneld}
\end{figure}

In order to illustrate the magnitude of the
transverse distortion effect, we consider a simple
model 
\bea
E_u(x,0,t)&=&\frac{1}{2}\kappa_u H_u(x,0,t)
\nonumber\\
E_d(x,0,t)&=& \kappa_d H_d(x,0,t),
\label{eq:Emodel}
\eea
where $H_{u}(x,0,t)=2H_d(x,0,t)$ is taken from Eq. 
(\ref{eq:log2}) and $\kappa_u$ and $\kappa_d$ are 
determined by making the approximation that only $u$ 
and $d$ quarks contribute to the nucleon's magnetic 
moment, i.e.
\bea
\kappa_p&=&\frac{2}{3}\kappa_u -
\frac{1}{3}\kappa_n= 1.793 
\nonumber\\
\kappa_n&=&\frac{2}{3}\kappa_d-\frac{1}{3}
\kappa_u= -1.913
\label{eq:kappa0}
\eea

\bea
\kappa_u&=&2\kappa_p +\kappa_n
= 1.673
\nonumber\\
\kappa_d&=&2\kappa_n+\kappa_p
= -2.033.
\label{eq:kappa}
\eea
With $\int_{-1}^1 dx H_d(x,0,0) = 1$ and
$\int_{-1}^1 dx H_u(x,0,0)=2$
this ansatz satisfies\\ 
$\int_{-1}^1 dx E_u(x,0,0) = \kappa_u$ and
$\int_{-1}^1 dx E_d(x,0,0) = \kappa_d$.
The resulting distributions in impact parameter 
space for both unpolarized and `transversely 
polarized' protons are depicted in Figs. 
\ref{fig:panelu} and \ref{fig:paneld} for $u$ and
$d$ quarks respectively.

The above Ansatz for $E_q(x,0,t)$ (\ref{eq:Emodel}),
together with our model for the impact parameter 
distribution of quarks in a longitudinally
polarized nucleon (\ref{eq:model}), therefore
yields for the distributions $q_X(x,{\bf b_\perp})$
in a `transversely polarized' nucleon
\bea
u_X(x,{\bf b_\perp})&=&
\left[1+\frac{\kappa_uS_xb_y}{8Ma(1-x)\ln
\frac{1}{x}}\right]
u(x)\frac{1}{4\pi a (1-x)\ln \frac{1}{x}}
e^{-\frac{ {\bf b}_\perp^2 }{4a(1-x)\ln \frac{1}{x}}}
\nonumber\\
d_X(x,{\bf b_\perp})&=&
\left[1+\frac{\kappa_dS_xb_y}{4Ma(1-x)\ln
\frac{1}{x}}\right]
d(x)\frac{1}{4\pi a (1-x)\ln \frac{1}{x}}
e^{-\frac{ {\bf b}_\perp^2 }{4a(1-x)\ln \frac{1}{x}}}
,
\eea
where $S_x=\pm 1$ for $|X\rangle = \frac{1}{\sqrt{2}}
\left(|\uparrow\rangle \pm |\downarrow 
\rangle\right)$ [Figs. \ref{fig:panelu} and
 \ref{fig:paneld}].
One can make several observations from
Figs. \ref{fig:panelu} and \ref{fig:paneld}.
First of all, the distortion for $u$ and
$d$ quarks is in opposite directions. In our model
this results from the simple fact that 
$\kappa_u$ and $\kappa_d$ have opposite signs.
However, since $\int_{-1}^1dx\int d^2{\bf b_\perp}
{\cal E}_q(x,{\bf b_\perp})=\kappa_q$ 
should always hold, one expects the fact
that $u$ and $d$ quarks get distorted in 
opposite directions for a `transversely polarized'
state to be a more general phenomenon.
Physically, one can understand the difference in
signs by replacing the rigid rotation of the sphere
in Fig. \ref{fig:fspin} by the intrinsic orbital
angular momentum of the quarks, which may be either
parallel or anti-parallel to the nucleon spin.

Secondly, one observes that the sideways distortion 
for $d$ quarks is about twice
as strong as for $u$ quarks, even though
$\kappa_u$ and $\kappa_d$ have about the same
magnitude. This is because the undistorted 
distribution $u(x,{\bf b_\perp})$ is twice
as large as $d(x,{\bf b_\perp})$, while
$\left|{\cal E}_d(x,{\bf b_\perp})\right|
\approx \left|{\cal E}_d(x,{\bf b_\perp})\right|$,
and therefore adding 
$\frac{1}{2M}\frac{\partial}{\partial b_y}
{\cal E}_q(x,{\bf b_\perp})$ results in a larger 
distortion for $d$ quarks than for $u$ quarks.
Finally one notes that the resulting quark 
distribution in this state
has a transverse (electric) dipole moment 
\be
d_y = \frac{2}{3} d_y^u - \frac{1}{3}d_y^d,
\ee
where
\bea
d_y^q &\equiv& \int_{-1}^1 dx \int d^2{\bf b_\perp}
b_y\, q_X(x,{\bf b_\perp})
= \frac{1}{2M} \int_{-1}^1 dx \int d^2{\bf b_\perp}
{\cal E}(x,{\bf b_\perp})\nonumber\\
&=&  \int_{-1}^1 dx E_q(x,0,0)
= \frac{\kappa_q}{2M},
\label{eq:dipole}
\eea
i.e., using Eq. (\ref{eq:kappa}), one finds
a rather large numerical value for the
transverse electric dipole moment due to 
this distortion
\be
\mu_y^{el.} = 0.17 \, e\, fm .
\label{eq:dy}
\ee
Of course, one could have also read off the
order of magnitude from Figs. 
\ref{fig:panelu}, \ref{fig:paneld}, but in
contrast to the model results presented in
the figures, Eq. (\ref{eq:dipole}) is a model
independent result. Furthermore, although the 
precise numerical value for the transverse 
electric dipole moment
(\ref{eq:dy}) relies on the approximation 
$\kappa_s\approx 0$, a model independent 
measure of the transverse distortion 
(relying only on isospin symmetry) is still
given by $d^u_y-d^d_y = \frac{\kappa_u-\kappa_d}{2M}
=\frac{\kappa_p-\kappa_n}{2M}
\approx 0.38 \, fm $.
Note also that the opposite signs of $\kappa_u$ and
$\kappa_d$ imply that the distributions of $u$ and $d$
quarks get distorted in opposite directions
(Figs. \ref{fig:panelu}, \ref{fig:paneld})
and therefore they contribute with the same sign
(i.e. coherently) to the transverse electric
dipole moment.
This transverse electric dipole moment
is perpendicular
to both the momentum of the nucleon (along 
$z$-axis) as well as the polarization (in
$x$-direction) and an electric dipole moment
pointing in a direction determined by
\be
{\vec \mu}^{el.} \sim {\vec S} \times {\vec P},
\ee
is thus consistent with both parity and time reversal
invariance. In fact, one can observe certain 
similarities with the geometry that arises in 
single spin assymetries \cite{ssa}.

\subsection{Implications for Transverse Hyperon 
Polarization}
The transverse distortion of impact parameter
dependent PDFs for states with transverse 
polarization should have many interesting 
implications for the production of hadrons with 
transverse polarization as one can see from the
the reactions $P+P\longrightarrow
Y+X$ or $P+\bar{P}\longrightarrow
Y+X$ with $Y\in \left\{ \Lambda,\Sigma,\Xi
\right\}$ at high energies.
We will use plausibility arguments to motivate a
rather simple general reaction mechanism and
explain the implications for the transverse 
polarization of produced hyperons.

First we will assume that most hyperons are not 
produced in central collisions, but instead in 
more peripheral interactions because at high energy
a central collision will most likely yield only an
unpolarized background. Secondly, we will assume
that the hyperon will be deflected in the direction
where the initial proton overlaps with the
anti-proton because most reaction mechanisms between
$p$ and $\bar{p}$ will give rise to a strong
attraction. Finally, we will assume that the 
$s\bar{s}$ pair is produced in that overlap region
and that the produced $s$ quarks form the final state
hyperon together with the `rest' of the initial proton.

\begin{figure}
\unitlength1.cm
\begin{picture}(10,9)(-1,8)
\includegraphics{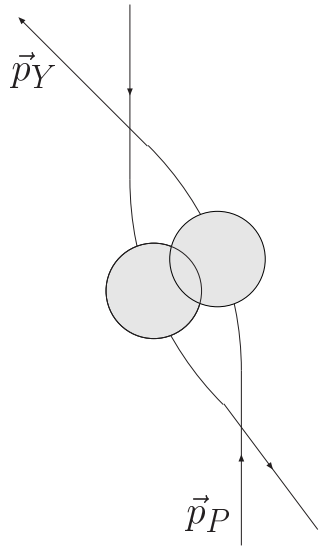}
\end{picture}
\caption{$P+P(\bar{P})\longrightarrow Y+\bar{Y}$
where the incoming $P$ (from bottom) 
is deflected to the left during the reaction.
The $s\bar{s}$ pair is assumed to be produced
roughly in the overlap region, i.e. on the left
`side' of the $Y$.}
\label{fig:collide}
\end{figure}

\begin{figure}
\unitlength1.cm
\begin{picture}(10,5)(1,13)
\includegraphics{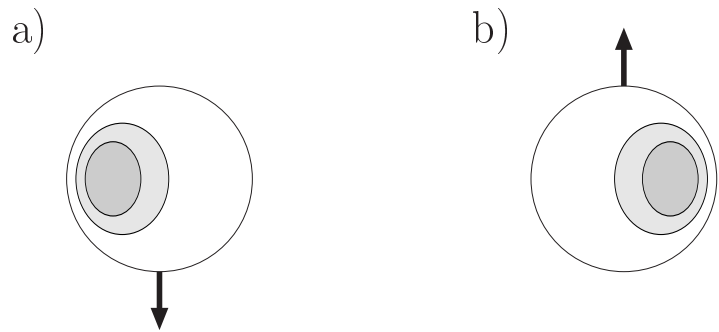}
\end{picture}
\caption{Schematic view of the transverse distortion
of the $s$ quark distribution (in grayscale)
in the transverse plane for a transversely
polarized hyperon with $\kappa_s^Y>0$. The view
is (from the rest frame) into the direction of 
motion (i.e. momentum into plane) for a hyperon 
that moves with a large momentum.
In the case of spin down (a), the $s$-quarks get distorted towards
the left, while the distortion is to the right
for the case of spin up (b).}
\label{fig:shift}
\end{figure}

These very simple assumptions,
together with the transverse distortion of the
strange quark cloud in a transversely polarized
hyperon will favor hyperons with a specific
transverse polarization: Consider first
the case where the
hyperon has been deflected to the left 
(Fig. \ref{fig:collide}), where
we look in the direction of the outgoing hyperon.
Based on our model assumptions above, the $s$ quark
has been produced on the left side of the
hyperon, which we assume has a positive
strange anomalous magnetic moment $\kappa^Y_s$ 
(the case $\kappa_s^Y<0$ yields the opposite effect).
If the final state hyperon is polarized `up'
(w.r.t. the reaction plane) then the $s$ quark 
distribution in the hyperon
is distorted to the right, i.e. away from the
reaction zone, while it is distorted to the left
(towards the reaction zone) if the polarization is
down (Fig. \ref{fig:shift}). 
Clearly the second possibility yields a
better overlap between the intermediate state and the
final state and we would therefore expect
a polarization `down' in this case.
The polarization is reversed when the hyperon is 
deflected to the right (because then the reaction 
zone is on its right side) and it is also reversed
when the sign of $\kappa_s^Y$ is reversed.
These simple considerations lead to
the prediction that the polarization direction
in this reaction is determined by
\be
{\vec P}_Y \sim - \kappa_s^Y {\vec p}_P
\times {\vec p}_Y.
\ee
Before we can compare this remarkably simple 
prediction to experimental data, we need to
determine $\kappa_s^Y$ for various hyperons.
Using $SU(3)$ flavor symmetry, one finds 
\be
\kappa_s^\Lambda &=& \kappa^p+\kappa_s^p
= 1.79 + \kappa_s^p\nonumber\\
\kappa_s^\Sigma &=& \kappa^p+2\kappa^n+
\kappa_s^p
= -2.03 + \kappa_s^p\nonumber\\
\kappa_s^\Xi &=& 2\kappa^p+\kappa^n +\kappa_s^p
= 1.67 + \kappa_s^p .
\label{eq:kappas}
\ee
Although the exact value for the strange magnetic
moment of the nucleon is not known, it is unlikely
to be on the same order as $\kappa^p$ or $\kappa^n$,
i.e. Eq. (\ref{eq:kappas}) tells us that the strange
anomalous magnetic moment of the $\Lambda$ and
the $\Xi$ hyperon is positive, while it is negative
for the $\Sigma$, yielding for the polarizations
\be
{\vec P}_\Lambda &\sim& -{\vec p}_P \times {\vec p}_Y
\nonumber\\
{\vec P}_\Sigma &\sim& +{\vec p}_P \times {\vec p}_Y
\nonumber\\
{\vec P}_\Xi &\sim& -{\vec p}_P \times {\vec p}_Y
\ee
which agrees with the experimentally observed 
pattern 
of polarizations ($P$ w.r.t. ${\vec p}_P\times
{\vec P}_Y$)
\be
0<P_{\Sigma^0}\approx
P_{\Sigma^-}\approx
P_{\Sigma^+}\approx -P_\Lambda
\approx -P_{\Xi^0}\approx -P_{\Xi^-}
\ee
in hyperon production reactions.\footnote{For
a recent discussion of polarization in such reactions,
see Ref. \cite{hyper} and references therein.}  

Although we illustrated the effect here in the
example of $P+P(\bar{P})\longrightarrow Y+X$, it
should be clear that the effect should also apply
to many other hyperon (as well as other hadron)
 production reactions, but
a complete discussion of this subject would
go beyond the scope of this section which should
serve only as a simple illustration of the
striking consequences that the transverse
distortion of the parton distribution in 
transversely polarized may have.

\section{Polarized Impact Parameter Dependent
Parton Distributions}
For a longitudinally polarized nucleon, 
i.e. when $\lambda=\lambda^\prime$, the
distribution of {\sl unpolarized} quarks
is of course independent of $\lambda$,
i.e.
\be
q(x,{\bf b_\perp}) \equiv 
\left\langle p^+,{\bf R}_\perp=0,\lambda
\right|
\hat{O}_q(x,{\bf b_\perp})
\left| p^+,{\bf R}_\perp=0,\lambda
\right\rangle , 
\label{eq:defl1}
\ee
regardless of $\lambda$ (this follows from
PT invariance). In this section
we are more
interested in impact parameter dependent
{\sl polarized} quark distributions, which
we {\sl define} as
\be
\Delta q(x,{\bf b_\perp}) \equiv 
\left\langle p^+,{\bf R}_\perp=0,\uparrow
\right|
\hat{O}_5(x,{\bf b_\perp})
\left| p^+,{\bf R}_\perp=0,\uparrow
\right\rangle , 
\label{eq:defpol},
\ee
where
\be
\hat{O}_5(x,{\bf b_\perp}) \equiv
\int \frac{dx^-}{4\pi}\bar{q}
\left(-\frac{x^-}{2},{\bf b_\perp} \right) 
\gamma^+ \gamma_5
q\left(\frac{x^-}{2},{\bf b_\perp}\right) 
e^{ixp^+x^-}.
\label{eq:bperp5}
\ee 
An equivalent 
definition with $\lambda = -\frac{1}{2}$ would
involve a minus sign on the r.h.s. of Eq. 
(\ref{eq:defpol}).
The physical significance of 
$\Delta q(x,{\bf b_\perp})$ becomes apparent
after we repeat the same steps as in
Eqs. (\ref{eq:step1}-\ref{eq:result1}), which
yields after some straightforward algebra
\be 
\Delta q(x,{\bf b_\perp}) 
=
 \int \frac{d^2{\bf \Delta}_\perp}{(2\pi)^2}  
\tilde{H}(x,-{\bf \Delta}_\perp^2) e^{-i{\bf b_\perp} \cdot
{\bf \Delta}_\perp},
\label{eq:resultpol}
\ee
i.e. by Fourier transform one can relate the
impact parameter dependent polarized quark
distribution $\Delta q(x,{\bf b_\perp})$ to
the polarized GPD $\tilde{H}(x,-{\bf \Delta}_\perp^2)$.

The impact parameter dependent polarized quark
distribution also satisfies
\be
\int d^2{\bf b_\perp} 
\Delta q(x,{\bf b_\perp}) =\Delta q(x),
\ee
as one might have expected. The probabilistic
interpretation of polarized impact parameter
dependent quark distributions is rather 
similar to the one of regular polarized quark
distribution: For this purpose, let 
us first define impact parameter dependent
parton distribution for parallel and anti-parallel
helicities $(x>0)$, 
\be
q_\uparrow(x,{\bf b_\perp}) \equiv 
\left\langle p^+,{\bf R}_\perp=0,\uparrow
\right|
\hat{O}_\uparrow(x,{\bf b_\perp})
\left| p^+,{\bf R}_\perp=0,\uparrow
\right\rangle , 
\label{eq:defpol1}
\\
q_\downarrow(x,{\bf b_\perp}) \equiv 
\left\langle p^+,{\bf R}_\perp=0,\uparrow
\right|
\hat{O}_\downarrow(x,{\bf b_\perp})
\left| p^+,{\bf R}_\perp=0,\uparrow
\right\rangle , 
\label{eq:defpol2}
\ee
with 
$\hat{O}_{\uparrow\downarrow}(x,{\bf b_\perp})
= \hat{O}(x,{\bf b_\perp})\pm
\hat{O}_5(x,{\bf b_\perp})$.
They provide a decomposition of the unpolarized
and polarized impact parameter dependent PDF
into helicity components, i.e.
\bea
q(x,{\bf b_\perp})&=&
q_\uparrow(x,{\bf b_\perp})+
q_\downarrow(x,{\bf b_\perp})
\quad \quad \quad \mbox{for}\,\,x>0
\nonumber\\
\Delta q(x,{\bf b_\perp})&=&
q_\uparrow(x,{\bf b_\perp})-
q_\downarrow(x,{\bf b_\perp})
\eea
and satisfy
\bea
\int d^2{\bf b_\perp} 
q_\uparrow(x,{\bf b_\perp})
&=& q_\uparrow (x) \nonumber\\
\int d^2{\bf b_\perp} 
q_\downarrow(x,{\bf b_\perp})
&=& q_\downarrow (x) 
\eea
Furthermore, like their forward counterparts,
they each have a probabilistic interpretation,
i.e. $q_\uparrow(x,{\bf b_\perp})\geq0$ and
$q_\downarrow(x,{\bf b_\perp})\geq0$, which
can also be expressed in the form 
\be
\left|q(x,{\bf b_\perp})\right| \geq 
\left| \Delta q(x,{\bf b_\perp})\right|.
\label{eq:ineqD}
\ee
Very similar relations can be obtained for 
antiquarks, where one can define $(x<0)$
\bea
\bar{q}(x,{\bf b_\perp}) &\equiv& -q(-x,{\bf b_\perp})
\nonumber\\
\Delta
\bar{q}(x,{\bf b_\perp}) &\equiv& 
\Delta
q(-x,{\bf b_\perp})
\eea
which can also be expressed in terms of (positive)
helicity dependent anti-quark distributions
$\bar{q}_{\uparrow \downarrow}(x,{\bf b_\perp})$
\bea
\bar{q}(x,{\bf b_\perp})&=&
\bar{q}_\uparrow (x,{\bf b_\perp})
+ \bar{q}_\downarrow (x,{\bf b_\perp})\nonumber\\
\Delta \bar{q}(x,{\bf b_\perp})&=&
\bar{q}_\uparrow (x,{\bf b_\perp})
- \bar{q}_\downarrow (x,{\bf b_\perp}).
\eea

\section{Polarized distributions with
nucleon helicity flip}
In Sec. \ref{sec:flip} we elucidated the physics
of the GPD $E_q$ by considering the (unpolarized)
quark distribution $q_{X} (x,{\bf b_\perp})$
in a target with $\perp$
polarization, and found that
$
q_{X} (x,{\bf b_\perp})=q(x,{\bf b_\perp})
- \frac{1}{2M}\frac{\partial}{\partial b_y}
{\cal E}_q(x,{\bf b_\perp}),
$
i.e. the unpolarized quark distribution gets distorted 
when the polarization is not in the longitudinal
direction. Naively one would expect that something
similar happens for polarized parton
distributions in the transverse plane.
However, as one can read off from Eq. (\ref{eq:gpd2}),
there is no contribution to the transition
amplitude from $\tilde{E}_q$ when $\Delta^+=0$.
As a result, one finds that
\be
\left\langle p^+,{\bf R_\perp}={\bf 0_\perp},\lambda
\left| \hat{O}_{5,q}(x,{\bf b_\perp}) \right|
p^+,{\bf R_\perp}={\bf 0_\perp},\lambda^\prime
\right\rangle = \delta_{\lambda \lambda^\prime}
\Delta q(x,{\bf b_\perp}),
\ee
without any contribution from $\tilde{E}_q$.
The physical interpretation of $\tilde{E}_q$ must
therefore be very different from the interpretation
of $E_q$. In particular, one should not expect
a simple probabilistic interpretation.

When the nucleon is not in a helicity eigenstate
but rather a superposition of helicity eigenstates,
e.g.
\be 
\left| p^+, {\bf R_\perp}={\bf 0_\perp}, \phi\right\rangle
\equiv \cos \phi
\left| p^+, {\bf R_\perp}={\bf 0_\perp}, \uparrow \right\rangle
+ \sin \phi 
\left| p^+, {\bf R_\perp}={\bf 0_\perp}, \downarrow\right\rangle
\ee
such that the mean helicity is $h=\cos 2\phi $ and the mean 
transverse spin is $S_x=\sin 2\phi$, then 
the polarized quark distribution 
$\Delta q_\phi(x,{\bf b_\perp})$
in the transverse
plane for such a state is obtained as a simple
rescalation from the helicity eigenstates, i.e.
\be
\Delta q_\phi(x,{\bf b_\perp}) \equiv 
\left\langle p^+, {\bf R_\perp}={\bf 0_\perp}, \phi\right|
\hat{O}_{5}(x,{\bf b_\perp})
\left| p^+, {\bf R_\perp}={\bf 0_\perp}, \phi\right\rangle
= h \Delta q (x,{\bf b_\perp}).
\label{eq:Deltaqh}
\ee
This is in contrast to the result for
unpolarized quark distributions in the same state
\be
q_\phi (x,{\bf b_\perp})=q(x,{\bf b_\perp})
- \frac{S_x}{2M}\frac{\partial}{\partial b_y}
{\cal E}_q(x,{\bf b_\perp}),
 \label{eq:qSx}
\ee
One can use these results to derive further
positivity constraints. For this purpose,
consider the distribution of quarks ($x>0$)
\bea
q_{\uparrow,\phi}(x,{\bf b_\perp}) &\equiv&
\left\langle p^+, {\bf R_\perp}={\bf 0_\perp}, \phi\right|
\hat{O}_{\uparrow}(x,{\bf b_\perp})
\left| p^+, {\bf R_\perp}={\bf 0_\perp}, \phi\right\rangle
\nonumber\\
&=&q_{\phi}(x,{\bf b_\perp}) + \Delta q_\phi (x,{\bf b_\perp})
\eea
with positive helicity in a state with such a
mixed helicity. Combining Eq. (\ref{eq:qSx})
with (\ref{eq:Deltaqh}) one thus obtains 
\be
q_{\uparrow,\phi}(x,{\bf b_\perp}) = q(x,{\bf b_\perp})
- \frac{S_x}{2M}\frac{\partial}{\partial b_y}
{\cal E}_q(x,{\bf b_\perp}) + h \Delta q (x,{\bf b_\perp})
\label{eq:quph}.
\ee
Following arguments very similar to the ones
in Section \ref{sec:impact}, one can easily convince oneself
that $q_{\uparrow,\phi}(x,{\bf b_\perp})$ should be
positive (for $x>0$). Since this should hold for arbitrary
$\phi$, one easily obtains
\be
\left| q(x,{\bf b_\perp}) \right|^2
\geq \left| \Delta q (x,{\bf b_\perp}) \right|^2
+ \left|\frac{1}{2M} \frac{\partial}{\partial b_y}
{\cal E}_q(x,{\bf b_\perp})\right|^2,
\label{eq:ineqEq}
\ee
which is stronger than Eqs. (\ref{eq:ineqD})
and \ref{eq:ineqE}. Note that Eq. (\ref{eq:ineqEq})
can also be expressed in the form
\be
\sqrt{\left|  q_\uparrow(x,{\bf b_\perp}) 
q_\downarrow (x,{\bf b_\perp}) \right|} \geq
\frac{1}{4M} \left|{\bf \nabla_{b_\perp}}
{\cal E}_q(x,{\bf b_\perp})\right| .
\ee
Using the known asymptotic behavior of 
$q_{\uparrow}$ and $q_\downarrow$ near
$x\rightarrow 1$ \cite{bbs}, one can
use the latter inequality to place an
upper bound on the asymptotic behavior of
${\bf \nabla_{b_\perp}}{\cal E}_q \sim (1-x)^4$.

\section{Nonzero Skewedness}
Deeply virtual Compton scattering experiments
always probe $\xi \neq 0$. Even though 
polynomiality conditions on the possible
$\xi$-dependence \cite{vdh} should facilitate 
the extrapolation of GPDs extracted from 
DVCS experiments to $\xi=0$, it would nevertheless
be very desirable to develop a better physical 
understanding of GPDs at $\xi\neq 0$.
In order to better appreciate the difficulties
in developing such an interpretation, let us start
out by summarizing the problem from an abstract
point of view. 
In general, GPDs have the physical interpretation
of a transition amplitude. 
A necessary condition for a density interpretation
is that the initial and final state are the same.
This is the case for the usual PDFs but it is not
for GPDs, since the latter are distinguished from
PDFs due to the fact that $p^\prime \neq p$.
In the case $\xi=0$,
i.e. when $\Delta^+\equiv p^{+\prime}-p^+$, 
a Fourier transform
of GPDs w.r.t. ${\bf \Delta_\perp}\equiv 
{\bf p_\perp^\prime} - {\bf p_\perp}$ diagonalizes
the transition amplitude and one can have a density
interpretation.

In the case $\Delta^+\neq 0$, a Fourier transform
w.r.t. ${\bf \Delta_\perp}$ is obviously not enough 
to render the initial and final state the same, because
even after Fourier transforming transverse momenta, 
the longitudinal momenta are still sharp --- and
different in the initial and final states.
One might be tempted to try using yet another
Fourier transform w.r.t. the longitudinal momentum
of the target, in order to completely diagonalize
GPDs. However, this cannot make sense because the
$x$-variable in GPDs already measures the 
longitudinal momentum of the active quark and 
Heisenberg's uncertainly relations therefore 
prohibit the simultaneous
measurement of the longitudinal position
of the active quark. Even if one would be willing
to give up on a precise measurement of $x$, the
limitations due to the Heisenberg inequality
would severely restrict the longitudinal
position space interpretation of GPDs. One can get
a rough estimate for the quantitative limitations of 
the resulting resolution by reminding oneself that 
the Compton wavelength
of a proton is about $1/3$ the size of the proton.
Therefore even if one measures the momentum of a 
parton only with an accuracy that is about $1/3$ of
the nucleons momentum, the resulting 
position space uncertainty would be of about the 
same scale as the size of the proton. Therefore, 
even though one can formally proceed and develop
a semiclassical mathematical formalism to extract 
a parton distribution in 3-dimensional position space
$q(x,b^-,{\bf b_\perp})$ from GPDs, the 
limitations due to the uncertainty principle would
render this kind of distribution meaningless for a 
proton target.

In Refs. \cite{mb1,mb2} a mixed representation
(longitudinal momentum/transverse position) was
used to simplify the physical interpretation of
GPDs with $\xi=0$ as a density. Even though we 
argued above that such a basis does not lead to
a probabilistic interpretation for $\xi\neq 0$,
it may nevertheless be useful in simplifying the
physical interpretation of GPDs with $\xi\neq 0$,
since such a basis may at least partially
diagonalize the degrees of freedom.

For this purpose, let us consider a basis of 
transversely localized basis states\cite{diehl2}
that are translated away from the origin
\be
\left| p^+,{\bf R_\perp}={\bf b_\perp},\lambda
\right\rangle \equiv {\cal N}
\int d^2{\bf p_\perp} e^{-i{\bf p_\perp}\cdot
{\bf b_\perp}}
\left| p^+,{\bf p_\perp},\lambda \right\rangle
.
\ee
In a recent paper, Diehl pointed out that
the Fourier transform of 
$H_q(x,\xi,-{\bf \Delta}_\perp^2)$ can be related in
a simple manner to matrix elements of the light-cone
correlator between such basis states\cite{diehl2}
\be
\int \frac{d^2 {\bf D_\perp}}{(2\pi)^2}
e^{-i{\bf D_\perp}\cdot {\bf b_\perp}}
H_q(x,\xi,{\bf \Delta}_\perp)&=&\label{eq:diehl2}\\
& & \!\!\!\!\!\!\!\!\!\!\!\!\!\!\!\!\!\!\!\!
 \frac{1+\xi^2}{(1-\xi^2)^{5/2}}
\left\langle p^{\prime +}, 
-\frac{\xi {\bf b_\perp}}{1-\xi}, \lambda \left|
{\cal O}_q(x,{\bf b_\perp})\right|p^+, 
\frac{\xi {\bf b_\perp}}{1+\xi},\lambda\right\rangle ,
\nonumber
\ee
where ${\bf D_\perp}(1-\xi^2) ={\bf \Delta}_\perp$.
Switching to an impact parameter representation yields
a transition amplitude that is still off diagonal in $p^+$,
which is not a surprise. However, on top of that
Eq. (\ref{eq:diehl2}) is also off-diagonal in the
impact parameter coordinate, which may be surprising
at first since the transition
amplitude in Eq. (\ref{eq:gpd}) neither changes the
transverse position of the spectators (which is trivial) 
nor does it change the transverse position of the active 
quark (since $ \hat{O}_q(x,{\bf b}_\perp) $ is diagonal
in transverse position).
Nevertheless, the transverse center of momentum changes
because the $p^+$ momentum of the active quark changes.
As a result, the transverse center of momentum changes
and therefore the distance of the active quark to the
transverse center of momentum changes.
In a sense the situation is comparable to the situation
of a decay in a nonrelativistic system when the mass
of the active quark changes and therefore the position 
of the center of mass changes, even if the positions of 
the individual partons remain the same in the transition
process. 

However, since the position of the transverse center of
mass of the spectators does {\it not} change in the
transition, it may be useful to switch to a basis,
where the transverse
position of the active quark is measured 
w.r.t. the center of momentum of the spectators, rather
than the center of momentum of the whole hadron,
because the transition amplitude would be diagonal
in that basis. Of course, it would still be off-diagonal
in the $p^+$ momentum.

\section{Summary}
Generalized parton distributions provide a light-cone
momentum decomposition for form factors in the sense
that they tell us how much quarks with a given
momentum fraction $x$ contribute to the form factor.

We considered a (proton)
state with a sharp $p^+$ momentum, which is
localized in the $\perp$ direction in the sense
that its {\sl transverse center of momentum}
${\bf R}_\perp \equiv \sum_i x_i {\bf r}_{\perp,i}$
is at the origin. Such a definition makes sense 
because there is a Galilean subgroup of transverse
boosts in the infinite momentum frame \cite{soper} 
and as a result one can separate the overall $\perp$ 
momentum from the internal dynamics. This is 
reminiscent of nonrelativistic dynamics, where one 
can work in the center of mass frame and where 
localizing the state by means of a wave packet 
corresponds to working in a frame where the center 
of mass is at the origin.

For such a state, we defined the notion
of {\sl impact parameter dependent parton distribution
functions}
$q(x,{\bf b_\perp})$, where ${\bf b_\perp}$ is
measured relative to the transverse center of 
momentum ${\bf R}_\perp$. The significance of
these impact parameter dependent PDFs 
$q(x,{\bf b_\perp})$ is that they are the 
Fourier transform of the $\xi=0$ GPD 
$H_q(x,0,-{\bf \Delta}_\perp^2)$. 
Moreover, $q(x,{\bf b_\perp})$ 
satisfies a positivity
constraint for $x>0$ and a negativity constraint for 
$x<0$, i.e. one can interpret $q(x,{\bf b_\perp})$ 
as a density.
\footnote{Of course, we have only shown that this 
probabilistic interpretation holds in the infinite
momentum frame (and with light cone gauge), but this
is no real drawback since that is anyways 
the only frame where even forward PDFs have a 
probabilistic interpretation.}
This aspect is very important because the 
probabilistic interpretation underscores the fact 
that impact parameter dependent PDFs have a physical 
significance --- above and beyond being 
the Fourier transforms of GPDs w.r.t. 
${\bf \Delta}_\perp$.

Conventional parton distributions contain no
information about the spatial distribution of 
partons. If one knows GPDs for $\xi=0$ one can 
simultaneously determine (by Fourier transform 
w.r.t. ${\bf \Delta}_\perp$) the longitudinal
momentum and transverse position of partons
in the target.\footnote{This result is {\sl not}
in contradiction with Heisenberg's uncertainty
principle since the information extracted from
GPDs is only a simultaneous determination of 
{\sl different} components of momentum and
position of partons.}
This is completely new information
and should provide us with new insights about 
the internal structure of hadrons.
For example, knowledge of GPDs allows to answer
questions like `how the spatial distribution of
partons in the nucleons varies with x' or `what
is the parton distribution at a given distance
from the center of momentum'.

One of the remarkable results is that there are
no relativistic corrections to this interpretation.
Formally, this is due to the Galilean subgroup
of transverse boosts in the infinite momentum 
frame. The only limitation in transverse 
resolution is due to the scale $\frac{1}{Q}$
with which one probes the target. The fact
that there are no relativistic corrections
to the transverse
position space interpretation of 
GPDs in the IMF implies
as a corollary that an interpretation of
the Fourier transform of form factors as
transverse position space distributions in the IMF
either.

$H_q(x,0,-{\bf \Delta_\perp^2})$ and 
$\tilde{H}_q(x,0,-{\bf \Delta_\perp^2})$ 
have the most simple physical interpretation as
Fourier transforms of unpolarized and polarized
impact parameter dependent PDFs respectively
for longitudinally polarized nucleons.
When the target is polarized in the transverse
direction, even the unpolarized quark distribution
changes. The resulting distortion of the
distribution of quarks in the transverse plane
is described by the Fourier transform of
the nucleon helicity flip distribution
$E_q(x,0,-{\bf \Delta_\perp^2})$.
The magnitude of the resulting transverse
flavor dipole moment can be related to the
anomalous magnetic moment for that flavor in
a model independent way.
For $\tilde{E}_q(x,0,-{\bf \Delta_\perp^2})$
we were unable to find a simple density
interpretation, since it does not contribute
to matrix elements with purely transverse
momentum transfer.

The transverse distortion of the impact parameter
dependent PDFs for transversely polarized targets
should have important consequences for
reactions involving quark production in peripheral 
scattering as we illustrated in the context
of transverse polarization of hyperons.

\section*{Acknowledgments}
This work was supported by a grant from DOE 
(FG03-95ER40965). 
It is a pleasure to thank Bob Jaffe, Xiangdong Ji,
and Andrei Belitsky for very useful discussions.
\appendix
\section{Galilean Subgroup of Transverse Boosts}
\label{sec:galilei}
Boost transformations in nonrelativistic quantum 
mechanics (NRQM)
\be
{\vec x}^\prime &=&{\vec x} + {\vec v}t
\nonumber\\
t^\prime&=&t
\ee
are purely kinematic because they leave the
quantization surface $t=0$ invariant. This property
has many important consequences. For example,
wavefunctions for a many body system in the rest 
frame and in a boosted frame are related by a simple
shift of (momentum) variables, e.g.
\be
\Psi_{\vec v}({\vec p}_1,{\vec p}_2,{\vec p}_3)
=\Psi_{\vec 0}({\vec p}_1-m_1{\vec v},
{\vec p}_2-m_2{\vec v},{\vec p}_3-m_3{\vec v}).
\label{eq:nrboost}
\ee
Furthermore, if the Hamiltonian is translationally
invariant, the dynamics of the {\sl center of mass}
\be
{\vec R}=\sum_i x_i {\vec r}_i,
\ee
with $x_i = m_i/M$ and $M=\sum_im_i$, separates from 
the intrinsic variables, making it possible to work 
in the center of mass frame.

One of the features that normally complicates the 
description of relativistic bound states is that
equal time hyperplanes are not invariant under
relativistic boosts
\be
{\bf x}_\perp^\prime &=&{\bf x}_\perp
\nonumber\\
z^\prime &=&
\gamma\left(z  + v t\right)
\nonumber\\
t^\prime&=&\gamma\left(t+\frac{v}{c^2}z\right),
\ee
with $\gamma^{-2}=1-\frac{v^2}{c^2}$. As a result,
boosts are in general a dynamical operation, the
generator of boost transformations contains
interactions and there exists no simple
generalization of Eq. (\ref{eq:nrboost}) to
a relativistic system quantized at equal times.
Furthermore, the notion of the center of mass has
no useful generalization in such an equal time
quantized relativistic framework.

One of the main advantages of the light-cone
(or infinite momentum) framework arise because
there is a subgroup of kinematical boosts among
the generators of the Poincare group \cite{soper}. 
To see this
let us start from the usual Poincar\'e algebra
\be
\left[P^\mu,P^\nu\right]&=&0 \label{eq:poincare}\\
\left[M^{\mu\nu},P^\rho\right] &=& i\left(
g^{\nu \rho}P^\mu - g^{\mu \rho}P^\nu\right)
\nonumber\\
\left[M^{\mu \nu}, M^{\rho \lambda}\right]
&=&i\left(g^{\mu \lambda}M^{\nu \rho}
+g^{\nu \rho}M^{\mu \lambda} 
-g^{\mu \rho}M^{\nu \lambda}
-g^{\nu \lambda}M^{\mu \rho}\right) 
\ee
where the generators of rotations and boosts are
respectively
$M_{ij}=\varepsilon_{ijk}J_k$ and $M_{i0}=K_i$.

We now introduce transverse `boost' 
operators\footnote{Strictly speaking
the physical meaning of these operators is a 
combination of boosts and rotations.}
\be
B_x &=& \frac{1}{\sqrt{2}}\left(K_x+J_y\right) 
\nonumber\\ 
B_y &=& \frac{1}{\sqrt{2}}\left(K_y-J_x\right) 
\label{eq:perpboost}
\ee
>From the Poincar\'e algebra (\ref{eq:poincare}),
it follows that these satisfy commutation relations
\be
\left[J_3,B_k\right]&=& i\varepsilon_{kl}B_l
\nonumber\\
\left[P_k,B_l\right] &=& -i\delta_{kl}P^+
\nonumber\\
\left[P^-,B_k\right] &=& -iP_k 
\nonumber\\
\left[P^+,B_k\right]&=& 0
\ee
with $k,l\in\{x,y\}$, $\varepsilon_{xy}=-
\varepsilon_{yx}=1$, and $\varepsilon_{xx}=
\varepsilon_{yy}=0$.
\footnote{Using these commutation relations, one 
easily 
verifies that $e^{i{\bf v}_\perp \cdot {\bf B_\perp}}
{\bf P}_\perp e^{-i{\bf v}_\perp \cdot {\bf B_\perp}}
= {\bf P}_\perp+P^+{\bf v}_\perp$, which justifies
the identification of ${\bf B}_\perp$ as a transverse
boost operator.}
Together with the well known commutation relations
\be
\left[J_z,P_k\right] &=& i\varepsilon_{kl}P_l
\\
\left[P^-,P_k\right] &=& \left[P^-,P^+\right]
=\left[P^-,J_z\right] = 0 \nonumber\\
\left[P^+,P_k\right] &=& \left[P^+,B_k\right]
=\left[P^+,J_z\right]=0 
\ee
these are the same commutation relations as
the commutation relations among the generators of 
the Galilei transformations for NRQM in the plane, 
provided we make the identifications
\be
P^-&\longrightarrow& \mbox{Hamiltonian}
\\
{\bf P}_\perp &\longrightarrow& \mbox{momentum in the
plane}
\nonumber\\
P^+ &\longrightarrow& \mbox{mass}
\nonumber\\
L_z&\longrightarrow& \mbox{rotations around $z$-axis}
\nonumber\\
{\bf B_\perp} &\longrightarrow&
\mbox{generator of boosts in the plane}
\nonumber .
\ee     
Because of this isomorphism between transverse boosts
in the infinite momentum frame and boosts in the
context of NRQM in the plane, many familiar results
from NRQM can be directly carried over to 
relativistic systems in the infinite momentum frame.

In order to construct a localized nucleon state, we 
start from so called infinite
momentum frame helicity states \cite{soper1}
$\left| p^+,{\bf p}_\perp,\lambda\right\rangle$.
They are defined by making use of a Wigner 
construction, where one starts from a massive
particle at rest with spin projection $\lambda$
along the $z$ axis and applies an appropriate
boost
\be
\left| p^+,{\bf p}_\perp,\lambda\right\rangle
= e^{-i{\bf v}_\perp \cdot {\bf B_\perp}}
\left| p^+,{\bf 0}_\perp,\lambda\right\rangle
= e^{-i{\bf v}_\perp \cdot {\bf B_\perp}}
e^{-i\omega K}
\left| M/\sqrt{2},{\bf 0}_\perp,\lambda\right\rangle
,
\ee
where ${\bf p}_\perp = p^+ {\bf v}_\perp$ and
$e^\omega = \sqrt{2}p^+/M$. 
These states are of well defined for any momentum
$p^\mu$. However, they are most useful to describe
particles that move with high velocity in the $z$
direction because, when viewed from the infinite
momentum frame these `light-cone helicity' states
also become eigenstates of the ordinary helicity
operator. For details we refer to Ref. 
\cite{soper1}. 
For our purposes the most important properties of 
these states are \cite{soper1}
\bea
e^{-i\phi J_z} \left|p^+,p^1,p^2,\lambda
\right\rangle
&=&
e^{-i\phi \lambda} \left|p^+,p^1\cos \phi 
-p^2\sin \phi,p^2\cos \phi + p^1 \sin \phi,\lambda
\right\rangle
\label{eq:prop1}\\
e^{-i{\bf v}_\perp \cdot {\bf B_\perp}}
\left| p^+,{\bf p}_\perp,\lambda\right\rangle
&=&
\left| p^+,{\bf p}_\perp+ p^+{\bf v}_\perp,
\lambda\right\rangle
\label{eq:prop2}.
\eea

>From these properties it is straightforward to show
that these states satisfy
\bea
\left| p^+,{\bf R}_\perp=0,\lambda
\right\rangle \equiv 
{\cal N} \int d^2 {\bf p}_\perp \left| 
p^+,{\bf p}_\perp,\lambda\right\rangle
\eea
are a simultaneous eigenstate of the longitudinal
(light-cone) momentum $P^+=P^0+P^3$, the total
angular momentum in the $z$ direction $J_z$, and
the transverse position operator ${\bf R}_\perp
= -\frac{1}{p^+} {\bf B_\perp}$, i.e.
\bea
\hat{P}^+
\left| p^+,{\bf R}_\perp=0,\lambda
\right\rangle
&=& p^+
\left| p^+,{\bf R}_\perp=0,\lambda
\right\rangle
\nonumber\\
{\bf R}_\perp
\left| p^+,{\bf R}_\perp=0,\lambda
\right\rangle &=&0 \nonumber\\
J_z\left| p^+,{\bf R}_\perp=0,\lambda
\right\rangle
&=&
\lambda
\left| p^+,{\bf R}_\perp=0,\lambda
\right\rangle .
\eea

In nonrelativistic quantum mechanics, boost
transformations shift the momentum of the $i^{th}$ 
particle by the amount
\be
{\vec p}_i \rightarrow {\vec p}_i + \Delta
{\vec v} m_i,
\ee
i.e. the generator of boosts is, up to the overall
mass of the system, given by the center of mass
operator
${\vec B}\equiv -\sum_i m_i {\vec r}_i = 
-M{\vec R}_{CM} $.

In the infinite momentum frame, ${\bf b_\perp}$,
the generator of $\perp$ boosts, has the physical 
meaning of the $\perp$ center of momentum times
the total momentum ${p^+}$ of the system.
There are several ways to see this result, which are
all worth mentioning since this illustrates the
physics of the $\perp$ center of momentum.
First this should by obvious due to the isomorphism
between boosts in nonrelativistic quantum mechanics
and $\perp$ boosts in the IMF provided one identifies
the nonrelativistic masses $m_i$ with the 
longitudinal momenta $k_i^+$, i.e. \cite{soper1}
\be
{\bf R_i} = - \frac{1}{P^+}{\bf B}_i.
\ee
Another way to arrive at the same conclusion is to
notice that $\Theta^{++}$, where $\Theta^{\mu \nu}$ 
is the energy momentum tensor, has the physical 
meaning of a light-cone momentum density. 
The field theoretic way of constructing 
the $\perp$ center of momentum is to weigh the 
$\perp$ position variable ${\bf x}_\perp$ with
the momentum density $\Theta^{++}$, yielding
\be
{\bf R}_\perp = \frac{1}{P^+}
\int dx^- d^2x_\perp \Theta^{++} 
x_\perp = -\frac{1}{P^+}{\bf B_\perp} ,
\label{eq:R=B}
\ee
which agrees with the result obtained by analogy with
nonrelativistic quantum mechanics. Furthermore
it shows how the transverse center of momentum 
operator is related to the generators of the
Poincare' group which are of course renormalization
group invariant.

The intuitive parton representation for 
${\bf R}_\perp$ in LF gauge is obtained
by expressing $\Theta^{++}$ in Eq. (\ref{eq:R=B})
in terms of light-cone creation and annihilation
operators and noticing that, after integrating
over $x^-$, only terms that are diagonal in Fock
space contribute, yielding
\be
{\bf R}_\perp = \frac{\sum_i k^+_i r_{\perp,i}}
{P^+_{total}} = \sum_i x_i r_{\perp,i}.
\ee
Again, this result should not be surprising, since 
the longitudinal momentum 
fractions $x_i \equiv k^+_i/P_{total}^+$ play a very 
similar role as the mass fractions $m_i/M_{total}$ 
in NRQM. 

The reason that ${\bf R}_\perp$ plays the role of
a reference point for impact parameter dependent
parton distributions can again be understood by
simple analogy with nonrelativistic quantum 
mechanics. We defined $q(x,{\bf b_\perp})$ starting
from the state
\be
|p^+,{\bf R_\perp}={\bf 0_\perp},\lambda
\rangle \equiv {\cal N} \int d^2p_\perp
|p^+,{\bf p}_\perp,\lambda \rangle.
\ee
>From the commutation relations it is clear that
${\bf b_\perp}$ is canonically conjugate to
${\bf p}_\perp$ and therefore this state satisfies
\be
{\bf b_\perp} | 
p^+,{\bf R_\perp}={\bf 0_\perp},\lambda
\rangle =0,
\label{eq:B=0}
\ee
just like $\hat {\vec r}\int d^3p |{\vec p}\rangle=0$
in NRQM. One can see this also directly by noticing
that 
\cite{soper1}
\be
e^{i{\bf v}_\perp \cdot {\bf B}_\perp} \left|
p^+,{\bf p}_\perp, \lambda \right\rangle
=\left|
p^+,{\bf p}_\perp +p^+{\bf v}_\perp,\lambda
\right\rangle
\ee
i.e.
\bea
e^{i{\bf v}_\perp \cdot {\bf B}_\perp} 
\int d^2p_\perp
\left|
p^+,{\bf p}_\perp, \lambda \right\rangle
&=&\int d^2p_\perp
\left|
p^+,{\bf p}_\perp +p^+{\bf v}_\perp,\lambda
\right\rangle
\nonumber\\
&=&
\int d^2p_\perp \left| p^+,{\bf p}_\perp, \lambda
\right\rangle
\ee
and therefore ${\bf B}_\perp \int d^2p_\perp
\left| p^+,{\bf p}_\perp,\lambda \right\rangle=0$.

Because of Eqs. (\ref{eq:R=B}) and 
(\ref{eq:B=0}) it is justified to say that 
${\bf b_\perp}$ in $q(x,{\bf b_\perp})$ is measured 
w.r.t. the transverse center of momentum 
${\bf R}_\perp$.

\section{Overlap Integrals of Light-Cone Wave 
Functions}
\label{sec:overlap}
The simple boost properties of LF wave functions
have also bee used to
construct convenient expression for GPDs in terms of 
overlap integrals \cite{diehl}. The Galilean 
invariance implies that LF wavefunctions in a frame 
with ${\Delta}_\perp\neq 0$ can be obtained from 
those in the ${\Delta}_\perp = 0$ frame by means of 
a simple shift of variables
\be
{p_i^+}&\longrightarrow &{p_i^+}^\prime = p_i^+
\nonumber\\
{\bf p}_{\perp,i}&\longrightarrow &
{\bf p}_{\perp,i}^\prime = {\bf p}_{\perp,i}
+ x_i {\bf \Delta}_\perp.
\label{eq:boost}
\ee
For example, for the N-particle Fock space amplitude 
$\Psi^N_{{\bf P}_\perp}$ for a state with overall
momentum ${\bf P}_\perp$, one finds
\bea
\Psi^N_{{\bf P}_\perp}\left(x_1,{\bf k}_{\perp,1},
x_2,{\bf k}_{\perp,2},...,x_N, {\bf k}_{\perp,N}
\right)
&=& \Psi^N_{{\bf 0}_\perp}
\left(x_1,{\bf k}_{\perp,1} -x_1
{\bf P}_\perp,
x_2,{\bf k}_{\perp,2}-x_2{\bf P}_\perp,...\right)
\nonumber\\
& & \quad \times\,
\delta\left({\bf P}_\perp-\sum_{i=1}^N {\bf k}_{\perp,i}
\right).
\label{eq:boost2}
\eea
Note that the momentum fractions
$x_i\equiv\frac{p^+_i}{\sum_i p_i^+}$ carried by the 
$i^{th}$ parton play a similar role here as the
mass fractions in the corresponding nonrelativistic
boost formula.

Using this very simple boost property one can
express typical overlap integrals that appear
within the context of form factors and GPDs in
terms of overlap integrals involving internal
momenta only (\ref{eq:exact}).
For purely transverse momentum transfers ($\xi=0$)
these overlap integrals take on a particularly simple
form in terms of LC wave functions
(Fock space amplitudes)\cite{diehl}
$\Psi_N(x,{\bf k}_\perp)$
\bea 
H_q(x,-\!{\bf \Delta}_\perp^2) &=&
\sum_N \!\sum_j\!\!
\int\!\! \left[dx\right]_N \!\!\int\!\! \left[d^2{\bf k}_\perp\right]_N\!
\delta(x-x_j) \Psi^*_N(x_i,{\bf k}_{\perp,i}^\prime,
\lambda_i)
\Psi_N(x_i,{\bf k}_{\perp,i},\lambda_i)
\nonumber\\
& &\label{eq:exact}
\eea
where 
${\bf k}_{\perp, i}^\prime = {\bf k}_{\perp, i} - x_i {\bf \Delta}_\perp$ for
$i\neq j$ and
${\bf k}_{\perp, j}^\prime = {\bf k}_{\perp, j} 
+(1-x_j) {\bf \Delta}_\perp$.
The helicities of the partons are labeled by 
$\lambda_i$ (In the following we will only
consider amplitudes that are diagonal in the
helicities and we will therefore omit the
helicity labels in all expressions in order to
simplify the notation).

Note that Eq. (\ref{eq:exact}) is very
similar to the expression for the form factor in the Drell-Yan frame, except
that the $x$ of the `active' quark is not integrated over, and it is 
exact if one knows the $\Psi_N$ for {\it all} Fock components.
The overlap integrals in terms of momentum space 
wavefunctions in Eq. (\ref{eq:exact}) are very much 
reminiscent of similar expressions for form factors 
of nonrelativistic systems. In order to explore this 
analogy further, we consider Fock space amplitudes 
$\tilde{\Psi}^N_{{\bf P}_\perp}$ in a `mixed' 
(longitudinal momentum and transverse position 
space) representation 
\be
\tilde{\Psi}^N_{{\bf P}_\perp}(x_i,{\bf r}_{\perp,i})
 \equiv
\int \left[d^2{\bf k}_{\perp}\right]
e^{i\sum_i {\bf r}_{\perp,i} {\bf k}_{\perp,i}} 
\Psi^N_{{\bf P}_\perp}(x_i,{\bf k}_{\perp,i}).
\ee
The fact that Fock space amplitudes depend only on
`relative' individual momenta 
$\hat{\bf k}_{\perp,i} \equiv {\bf k}_{\perp,i} -
x_i {\bf P}_\perp$, implies that the position space
wavefunction (Fock space amplitude) $\tilde{\Psi}^N$
depends on the overall transverse momentum
${\bf P}_\perp$ only through an overall phase,
i.e. it can be expressed in the form
\be
\tilde{\Psi}^N_{{\bf P}_\perp}(x_i, 
{\bf r}_{\perp,i}) &=&
\int \left[d^2{\bf k}_{\perp}\right]_N 
e^{i\sum_i{\bf k}_{\perp,i}{\bf r}_{\perp,i}}
\Psi_{{\bf P}_\perp}(x_i,{\bf k}_{\perp,i})
\nonumber\\
&=& e^{i\sum_i x_i{\bf P}_\perp {\bf r}_{\perp,i}}
\int \left[d^2\hat{\bf k}_{\perp}\right]_N 
e^{i\hat{\bf k}_{\perp,i}{\bf r}_{\perp,i}}
\Psi_{{\bf 0}_\perp}(x_i,{\bf k}_{\perp,i})
\nonumber\\
&=& e^{i{\bf P}_\perp {\bf R}_\perp}
\tilde{\Psi}^N_{rel.}(x_i, \hat{\bf r}_{\perp,i}),
\label{eq:relative}
\ee
where  
\be
{\bf R}_\perp &\equiv& \sum_i x_i 
{\bf r}_{\perp,i} 
\label{eq:perpcm}
\ee
is the {\sl transverse center of momentum}
and the $\hat{\bf r}_i$ are some `relative'
coordinates, e.g.
\be
\hat{\bf r}_{\perp,i}&\equiv& {\bf r}_{\perp,i}
- {\bf R}_\perp .
\ee
In order to illustrate the physics consequences of
this decoupling of the overall transverse momentum
for $H_q(x,t)$,
let us consider the specific cases $N=2$ and $N=3$.
\footnote{We may consider each Fock component $N$
separately since $H_q(x,t)$ is  diagonal in Fock 
space!}
In a two particle system
\bea 
H_q(x,-\!{\bf \Delta}_\perp^2) = 
\int\!\! d^2{\bf k_\perp}
\psi^*_{\bf \Delta_\perp}\!
(x,{\bf k_\perp}\!+\!{\bf \Delta_\perp})
\psi_{\bf 0_\perp}\!(x,{\bf k}_\perp) 
,
\ee
and using (\ref{eq:relative})
\be
\psi_{\bf 0_\perp}(x,{\bf k}_\perp) &=&
{\cal N} \!\!\int \!\!d^2{\bf r}_{\perp,1}
d^2{\bf r}_{\perp,2} e^{-i{\bf k}_\perp
({\bf r}_{\perp,1}-{\bf r}_{\perp,2})}
\tilde{\psi}(x,{\bf r}_\perp)\\
\psi_{\bf \Delta_\perp}(x,{\bf k}_\perp
+{\bf \Delta}_\perp) &=& {\cal N}
e^{i{\bf \Delta}_\perp {\bf R}_\perp}
\!\!\int \!\!d^2{\bf r}_{\perp,1}
d^2{\bf r}_{\perp,2} 
e^{-i({\bf k}_\perp+{\bf \Delta}_\perp)
{\bf r}_{\perp,1}}
e^{i{\bf k}_\perp {\bf r}_{\perp,2}}
\tilde{\psi}(x,{\bf r}_\perp),\nonumber
\ee
where ${\cal N}$ is some normalization constant
such that ${\sl N}^2\int d^2{\bf R}_\perp=1$,
one straightforwardly finds
\be
H_q(x,-{\bf \Delta}^2) = \int d^2{\bf r}_\perp
e^{i({\bf r}_{\perp,1}-{\bf R}_\perp)\cdot 
{\bf \Delta_\perp}}
\left|\tilde{\psi}(x,{\bf r}_\perp)\right|^2.
\label{eq:psi2}
\ee
Eq. (\ref{eq:psi2}) makes it clear that 
$H_q(x,-{\bf \Delta}^2_\perp)$ is the Fourier 
transform
of a density. Furthermore, one can easily read off
that the relevant transverse position scale is
{\sl not} set by the separation 
${\bf r}_\perp\equiv{\bf r}_{\perp,1}
-{\bf r}_{\perp,2}$
between the quark and the antiquark but by
the separation of the active quark from the
transverse center of momentum (\ref{eq:perpcm}),
which in the center of momentum frame coincides
with the impact parameter of the active quark
\be
{\bf b}_{\perp,1} \equiv {\bf r}_{\perp,1}
-{\bf R}_\perp = (1-x){\bf r}_\perp  .
\ee
Starting from the light-cone wave functions,
we have thus verified the interpretation of the 
generalized parton distribution 
$H_q(x,-{\bf \Delta}^2)$ as the Fourier transform of 
the probability distribution for the active quark 
with respect to its impact parameter in the case 
of a two particle system.

For a three particle system, we start from
\bea 
H_q(x,-\!{\bf \Delta}_\perp^2) = 
\int\!\! d^2{\bf k}_{\perp,1} 
\!\!\int\!\! d^2{\bf k}_{\perp,2}
\!\!\int \!\! dy
\psi^*_{\bf \Delta_\perp}\!
(x,{\bf k}_{\perp,1}\!+\!{\bf \Delta_\perp},
y,{\bf k}_{\perp,2})
\psi_{\bf 0_\perp}\!(x,{\bf k}_{\perp,1},
y,{\bf k}_{\perp,2}), \!\!\!\!\!\!\!\!\!\nonumber\\ 
\label{eq:H3}
\ee
and use (\ref{eq:relative})
\bea
& & \!\!\!
\psi_{\bf 0_\perp}(x,{\bf k}_{\perp,1},
y,{\bf k}_{\perp,2}) = \label{eq:boost3}\\
& &
{\cal N}\!\! \int\!\! d^2{\bf r}_{\perp,1}
d^2{\bf r}_{\perp,2} d^2{\bf r}_{\perp,3}
e^{-i{\bf k}_{\perp,1}
({\bf r}_{\perp,1}-{\bf r}_{\perp,3})}
e^{-i{\bf k}_{\perp,2}
({\bf r}_{\perp,2}-{\bf r}_{\perp,3})}
\tilde{\psi}(x,\hat{\bf r}_{\perp,1},
y,\hat{\bf r}_{\perp,2})\nonumber\\
& &\!\!\!
\psi_{\bf \Delta_\perp}(x,{\bf k}_{\perp,1}
\!+\!{\bf \Delta}_\perp,y,{\bf k}_{\perp,2}) =
\nonumber\\ 
& & {\cal N}
e^{i{\bf \Delta}_\perp {\bf R}_\perp}\!\!
\int \!\!d^2{\bf r}_{\perp,1}
d^2{\bf r}_{\perp,2} d^2{\bf r}_{\perp,3}
e^{-i({\bf k}_{\perp,1}+{\bf \Delta}_\perp)
{\bf r}_{\perp,1}}
e^{-i{\bf k}_{\perp,2} {\bf r}_{\perp,2}}
e^{i({\bf k}_{\perp,1}+{\bf k}_{\perp,2}) 
{\bf r}_{\perp,3}}\nonumber\\
& & \quad \quad \quad \quad \quad
\times \tilde{\psi}(x,\hat{\bf r}_{\perp,1},
y,\hat{\bf r}_{\perp,2}),\nonumber
\ee
where ${\cal N}$ is again  some normalization 
constant. Upon inserting (\ref{eq:boost3}) into
(\ref{eq:H3}), one finds after some straightforward
algebra
\be
H_q(x,-\!{\bf \Delta}_\perp^2) &=& 
{\cal N}e^{i{\bf \Delta}_\perp {\bf R}_\perp}\!\!
 \int\!\! d^2{\bf r}_{\perp,1}
d^2{\bf r}_{\perp,2} d^2{\bf r}_{\perp,3}
e^{i{\bf \Delta}_\perp {\bf r}_{\perp,1}}
\left| \tilde{\psi}(x,\hat{\bf r}_{\perp,1},
y,\hat{\bf r}_{\perp,2})\right|^2,\nonumber\\
&=&
\int\!\! d^2{\bf r}_{\perp,1}
d^2{\bf r}_{\perp,2} 
e^{i{\bf \Delta}_\perp ({\bf r}_{\perp,1}-
{\bf R}_\perp )}
\left| \tilde{\psi}(x,\hat{\bf r}_{\perp,1},
y,\hat{\bf r}_{\perp,2})\right|^2,
\ee
i.e. again we confirm the physical interpretation
of $H_q(x,0,-{\bf \Delta}_\perp^2)$ as the Fourier 
transform of a density.

The generalization of this light-cone wave function 
based derivation of our general result to arbitrary
numbers of partons is straightforward and is left as
an exercise for the reader.

\end{document}